\documentclass{aastex62}

\usepackage{natbib}

\graphicspath{{./}{figures/}}

\accepted{AJ}

\shorttitle{The First Circumbinary Planet Discovered by TESS}
\shortauthors{Kostov et al.}

\newcommand{\eleanor}{\texttt{eleanor}}

\newcommand{\tess}{{TESS}}
\newcommand{\target}{TOI-1338}
\newcommand{\lightkurve}{\texttt{Lightkurve}}

\begin{document}

\title{TOI-1338: TESS' First Transiting Circumbinary Planet}

\correspondingauthor{Veselin Kostov}
\email{veselin.b.kostov@nasa.gov}

\author[0000-0001-9786-1031]{Veselin~B.~Kostov}
\affiliation{NASA Goddard Space Flight Center, 8800 Greenbelt Road, Greenbelt, MD 20771, USA}
\affiliation{SETI Institute, 189 Bernardo Ave, Suite 200, Mountain View, CA 94043, USA}
\author[0000-0001-9647-2886]{Jerome A. Orosz}
\affil{Department of Astronomy, San Diego State University, 5500 Campanile Drive, San Diego, CA 92182, USA}
\author[0000-0002-9464-8101]{Adina~D.~Feinstein}
\altaffiliation{NSF Graduate Research Fellow}
\affil{Department of Astronomy and Astrophysics, University of
Chicago, 5640 S. Ellis Ave, Chicago, IL 60637, USA}
\author[0000-0003-2381-5301]{William F. Welsh}
\affil{Department of Astronomy, San Diego State University, 5500 Campanile Drive, San Diego, CA 92182, USA}
\author{Wolf Cukier}
\affil{Scarsdale High School, 1057 Post Rd, Scarsdale, NY 10583, USA}
\author{Nader Haghighipour}
\affil{Institute for Astronomy, University of Hawaii-Manoa, Honolulu, HI 96822, USA}
\author[0000-0002-9644-8330]{Billy Quarles}
\affil{Center for Relativistic Astrophysics, School of Physics, 
Georgia Institute of Technology, 
Atlanta, GA 30332, USA}
\author[0000-0002-7595-6360]{David~V.~Martin}
\altaffiliation{Fellow of the Swiss National Science Foundation}
\affil{Department of Astronomy and Astrophysics, University of
Chicago, 5640 S. Ellis Ave, Chicago, IL 60637, USA}
\author[0000-0001-7516-8308]{Benjamin~T.~Montet}
\affil{School of Physics, University of New South Wales, Sydney NSW 2052, Australia}
\affil{Department of Astronomy and Astrophysics, University of
Chicago, 5640 S. Ellis Ave, Chicago, IL 60637, USA}
\author{Guillermo Torres}
\affil{Center for Astrophysics $\vert$ Harvard \& Smithsonian, 60 Garden Street, Cambridge, MA 02138, USA}
\author[0000-0002-5510-8751]{Amaury H.M.J. Triaud}
\affil{School of Physics \& Astronomy, University of Birmingham, Edgbaston, Birmingham B15 2TT, United Kingdom}
\author[0000-0001-7139-2724]{Thomas~Barclay}
\affiliation{NASA Goddard Space Flight Center, 8800 Greenbelt Road, Greenbelt, MD 20771, USA}
\affiliation{University of Maryland, Baltimore County, 1000 Hilltop Cir,
Baltimore, MD 21250, USA}
\author{Patricia Boyd}
\affiliation{NASA Goddard Space Flight Center, 8800 Greenbelt Road, Greenbelt, MD 20771, USA}
\author{Cesar Briceno}
\affil{Cerro Tololo Inter-American Observatory, CTIO/AURA Inc., 
Casilla 603, La Serena, Chile}
\author{Andrew Collier Cameron}
\affil{Centre for Exoplanet Science, SUPA, School of Physics and Astronomy, University of St Andrews, St Andrews KY16 9SS, United Kingdom}
\author{Alexandre C.M. Correia}
\affil{CFisUC, Department of Physics, University of Coimbra, 3004-516 Coimbra, Portugal}
\author[0000-0002-0388-8004]{Emily A. Gilbert}
\affiliation{Department of Astronomy and Astrophysics, University of
Chicago, 5640 S. Ellis Ave, Chicago, IL 60637, USA}
\affiliation{The Adler Planetarium, 1300 South Lakeshore Drive, Chicago, IL 60605, USA}
\affiliation{NASA Goddard Space Flight Center, Greenbelt, MD 20771, USA}
\affiliation{GSFC Sellers Exoplanet Environments Collaboration}
\author[0000-0002-4259-0155]{Samuel Gill}
\affil{Department of Physics, University of Warwick, Gibbet Hill Road, Coventry CV4 7AL, United Kingdom}
\author{Micha\"el Gillon}
\affil{Astrobiology Research Unit, Universit\'e de Li\`ege, 19C All\'ee du 6 Ao\^ut, 4000 Li\`ege, Belgium}
\author{Jacob Haqq-Misra}
\affil{Blue Marble Space Institute of Science, Seattle, Washington, USA}
\author{Coel Hellier}
\affil{Astrophysics Group, Keele University, Staffordshire, ST5 5BG, United Kingdom}
\author{Courtney Dressing}
\affil{Department of Astronomy, University of California at Berkeley Berkeley, CA 94720, USA}
\author{Daniel C. Fabrycky}
\affil{Department of Astronomy and Astrophysics, University of
Chicago, 5640 S. Ellis Ave, Chicago, IL 60637, USA}
\author{Gabor Furesz}
\affil{Department of Physics and Kavli Institute for Astrophysics and Space Research, Massachusetts Institute of Technology, Cambridge, MA 02139, USA}
\author{Jon Jenkins}
\affil{NASA Ames Research Center, Moffett Field, CA, 94035, USA}
\author{Stephen R. Kane}
\affil{Department of Earth and Planetary Sciences, University of California, Riverside, CA 92521, USA}
\author{Ravi Kopparapu}
\affiliation{NASA Goddard Space Flight Center, 8800 Greenbelt Road, Greenbelt, MD 20771, USA}
\author[0000-0001-9419-3736]{Vedad Kunovac Hod\v{z}i\'c}
\affil{School of Physics \& Astronomy, University of Birmingham, Edgbaston, Birmingham B15 2TT, United Kingdom}
\author[0000-0001-9911-7388]{David W. Latham}
\affil{Center for Astrophysics $\vert$ Harvard \& Smithsonian, 60 Garden Street, Cambridge, MA 02138, USA}
\author{Nicholas Law}
\affil{Department of Physics and Astronomy, University of North Carolina
at Chapel Hill, Chapel Hill, NC 27599-3255, USA}
\author{Alan M. Levine}
\affil{Department of Physics and Kavli Institute for Astrophysics and Space Research, Massachusetts Institute of Technology, Cambridge, MA 02139, USA}
\author[0000-0001-8308-0808]{Gongjie Li}
\affil{Center for Relativistic Astrophysics, School of Physics, 
Georgia Institute of Technology, 
Atlanta, GA 30332, USA}
\author{Chris Lintott}
\affil{Oxford University, United Kingdom}
\author[0000-0001-6513-1659]{Jack~J.~Lissauer}
\affil{NASA Ames Research Center, Moffett Field, CA 94035, USA}
\author[0000-0003-3654-1602]{Andrew~W.~Mann}
\affil{Department of Physics and Astronomy, University of North Carolina at Chapel Hill, Chapel Hill, NC 27599, USA}
\author{Tsevi Mazeh}
\affil{Department of Astronomy and Astrophysics, Tel Aviv University, 69978 Tel Aviv, Israel}
\author{Rosemary Mardling}
\affil{School of Physics \& Astronomy, Monash University, Victoria, 3800, Australia}
\author{Pierre F.L. Maxted}
\affil{Astrophysics Group, Keele University, Staffordshire, ST5 5BG, United Kingdom}
\author{Nora Eisner}
\affil{Oxford University, United Kingdom}
\author{Francesco Pepe}
\affil{Observatoire Astronomique de l'UniversitÃÂ© de Gen\'ÃÂšve, 51 chemin des maillettes, CH-1290 Sauverny, Switzerland}
\author{Joshua Pepper}
\affil{Department of Physics, Lehigh University, 16 Memorial Drive East, Bethlehem, PA 18015, USA}
\affil{Department of Physics and Kavli Institute for Astrophysics and Space Research, Massachusetts Institute of Technology, Cambridge, MA 02139, USA}
\author{Don Pollacco}
\affil{Department of Physics, University of Warwick, Gibbet Hill Road, Coventry CV4 7AL, United Kingdom}
\author{Samuel N. Quinn}
\affiliation{Center for Astrophysics \textbar \ Harvard \& Smithsonian, 60 Garden Street, Cambridge, MA 02138, USA}
\author{Elisa V. Quintana}
\affiliation{NASA Goddard Space Flight Center, 8800 Greenbelt Road, Greenbelt, MD 20771, USA}
\author{Jason~F.~Rowe}
\affil{Bishops University, 2600 College St, Sherbrooke, QC J1M 1Z7, Canada}
\author{George Ricker}
\affil{Department of Physics and Kavli Institute for Astrophysics and Space Research, Massachusetts Institute of Technology, Cambridge, MA 02139, USA}
\author{Mark E. Rose}
\affil{NASA Ames Research Center, Moffett Field, CA 94035, USA}
\author[0000-0002-6892-6948]{S.~Seager}
\affiliation{Department of Physics and Kavli Institute for Astrophysics and Space Research, Massachusetts Institute of Technology, Cambridge, MA 02139, USA}
\affiliation{Department of Earth, Atmospheric and Planetary Sciences, Massachusetts Institute of Technology, Cambridge, MA 02139, USA}
\affiliation{Department of Aeronautics and Astronautics, MIT, 77 Massachusetts Avenue, Cambridge, MA 02139, USA}
\author{Alexandre Santerne}
\affil{Aix Marseille Univ, CNRS, CNES, LAM, Marseille}
\author{Damien S\'egransan}
\affil{Observatoire Astronomique de l'UniversitÃÂ© de Gen\'ÃÂšve, 51 chemin des maillettes, CH-1290 Sauverny, Switzerland}
\author[0000-0001-5504-9512]{Donald R. Short}
\affil{Department of Astronomy, San Diego State University, 5500 Campanile Drive, San Diego, CA 92182, USA}
\author{Jeffrey C. Smith}
\affil{SETI Institute/NASA Ames Research Center, 189 Bernardo Ave, Suite 200, Mountain View, CA, 94043, US}
\author[0000-0002-7608-8905]{Matthew R. Standing}
\affil{School of Physics \& Astronomy, University of Birmingham, Edgbaston, Birmingham B15 2TT, United Kingdom}
\author{Andrei Tokovinin}
\affil{Cerro Tololo Inter-American Observatory, CTIO/AURA Inc., 
Casilla 603, La Serena, Chile}
\author{Trifon Trifonov}
\affil{Max Planck Institut fuer Astronomie, Heidelberg, Germany}
\author{Oliver Turner}
\affil{Observatoire Astronomique de l'UniversitÃÂ© de Gen\'ÃÂšve, 51 chemin des maillettes, CH-1290 Sauverny, Switzerland}
\author{Joseph D. Twicken}
\affil{SETI Institute, 189 Bernardo Ave, Suite 200, Mountain View, CA 94043, USA}
\affil{NASA Ames Research Center, Moffett Field, CA, 94035, USA}
\author{St\'ephane Udry}
\affil{Observatoire Astronomique de l'UniversitÃÂ© de Gen\'ÃÂšve, 51 chemin des maillettes, CH-1290 Sauverny, Switzerland}
\author{Roland Vanderspek}
\affil{Department of Physics and Kavli Institute for Astrophysics and Space Research, Massachusetts Institute of Technology, Cambridge, MA 02139, USA}
\author[0000-0002-4265-047X]{Joshua N. Winn}
\affil{Department of Astrophysical Sciences, Princeton University, Princeton, NJ 08544, USA}
\author{Eric T. Wolf}
\affil{Department of Atmospheric and Oceanic Sciences, Laboratory for Atmospheric and Space Physics, University of
Colorado Boulder, Boulder, Colorado, USA, USA}
\author{Carl Ziegler}
\affil{Dunlap Institute for Astronomy and Astrophysics, University of Toronto, 50 St. George Street, Toronto, Ontario M5S 3H4, Canada}
\author{Peter Ansorge}
\affil{Planet Hunters TESS}
\author{Frank Barnet}
\affil{Planet Hunters TESS}
\author{Joel Bergeron}
\affil{Planet Hunters TESS}
\author{Marc Huten}
\affil{Planet Hunters TESS}
\author{Giuseppe Pappa}
\affil{Planet Hunters TESS}
\author{Timo van der Straeten}
\affil{Planet Hunters TESS}

\begin{abstract}

We report the detection of the first circumbinary planet found by \tess. The target, a known eclipsing binary, was observed in sectors 1 through 12 at 30-minute cadence and in sectors 4 through 12 at two-minute cadence. It consists of two stars with masses of $1.1\,M_\odot$ and $0.3\,M_\odot$ on a slightly eccentric (0.16), 14.6-day orbit, producing prominent primary eclipses and shallow secondary eclipses. The planet has a radius of $\sim 6.9\, R_{\oplus}$ and was observed to make three transits across the primary star of roughly equal depths (${\sim 0.2\%}$) but different durations---a common signature of transiting circumbinary planets. Its orbit is nearly circular ($e\approx 0.09$) with an orbital period of $95.2$ days. The orbital planes of the binary and the planet are aligned to within ${\sim 1^{\circ}}$. To obtain a complete solution for the system, we combined the \tess\ photometry with existing ground-based radial-velocity observations in a numerical photometric-dynamical model. The system demonstrates the discovery potential of \tess\ for circumbinary planets, and provides further understanding of the formation and evolution of planets orbiting close binary stars. 

\end{abstract}


\section{Introduction\label{sec:intro}}

One of the most exciting breakthroughs from the {\it Kepler} mission was the discovery of circumbinary planets (CBPs). Four years of continuous observations of several thousand eclipsing binary stars 
\citep[EBs,][]{Prsa2011,Slawson2011,Kirk2016},
led to the discovery of 13 transiting CBPs orbiting 11 {\it Kepler} EBs 
\citep[][]{Doyle2011,Welsh2012,Orosz2012a,Orosz2012,Schwamb2013,Kostov2013,Kostov2014,Welsh2015,Kostov2016,Orosz2019,Socia2020}.
These discoveries spanned a number of firsts---e.g.~the first transiting CBP, the first CBP in the Habitable Zone (HZ), the first CBP in a quadruple star system, and the first transiting multiplanet CBP system. In addition to opening a new chapter in studies of extrasolar planets, Kepler's CBPs have confirmed theoretical predictions that planet formation in circumbinary configurations is a robust process, and suggest that many such planetary systems must exist \citep[e.g.\ ][]{Pierens2013,Kley2015}. Indeed, recent studies argue that the occurrence rate of giant, Kepler-like CBPs is comparable to that of giant planets in single-star systems \citep[$\sim 10\%;$][]{Armstrong2014,Martin2014,Li2016,Martin2019}. 

As exciting as the CBPs discovered from Kepler are, however, the present sample is small, likely hindered by observational biases, and leaves a vast gap in our understanding of this new class of worlds. This is not unlike the state of exoplanet science 20 years ago, when only a handful of hot Jupiter exoplanets were known. Pressing questions remain regarding the formation and migration efficiency of CBPs, their orbital architectures and occurrence rates, the formation, evolution and population characteristics of their host binary stars. Addressing these questions requires more CBP discoveries---which require continuous observations of a large number of EBs for prolonged periods of time. NASA's Transiting Exoplanet Survey Satellite \citep[\tess,][]{Ricker2015} will assist CBP discovery by observing roughly half a million EBs continuously for timespans between one month and one year for the nominal mission (Sullivan et al. 2015). This motivated us to continue our search for transiting CBPs by examining light curves of EBs observed by TESS.

Here we report the discovery of the first CBP from 
\tess---TOI-1338, a Saturn-size planet orbiting the known eclipsing binary star EBLM J0608-59\footnote{The target also has the designations TIC 260128333, TYC 8533-00950-1, Gaia DR2 5494443978353833088. Its Right Ascension and Declination are 06:08:31.97 and 
-59:32:28.08, respectively. It has \tess\ magnitude $T = 11.45\pm0.02$ mag, 
and $V = 11.72\pm0.02$ mag.} approximately every 95 days. At the time of this writing, this is the longest-period confirmed planet discovered by \tess. Below we present the details of our discovery, and discuss some of the characteristics of this newly-found CBP that allow us to place its discovery in a broader context. 

This paper is organized as follows. In Section \ref{sec:Discover} we briefly outline the discovery of the system, and describe the \tess\ data for the target star and the detection of the CBP transits. In Section \ref{sec:complementary} we present the complementary observations and data analysis; Section \ref{sec:ELC} outlines the photometric-dynamical analysis of the system. Section \ref{sec:discussion} presents a discussion of the results. We draw our conclusions in Section \ref{sec:end}. 

\section{Discovery\label{sec:Discover}}

\subsection{\tess\ Mission\label{subsec:TESS}}

The primary goal of the
\tess\ mission is to identify transiting planets around nearby bright stars that are amenable to follow-up characterization. \tess\ will observe about ${85\%}$ of the sky during its two-year primary mission \citep{Ricker2015}, using four cameras that provide a ${\rm 24^\circ \times 96^\circ}$ field-of-view (FOV); a sector is a ~${\rm \sim}$month-long observation of a single FOV. Most of the stars in the Full-Frame Images (FFIs) will be observed at 30-minute cadence, and 200,000 pre-selected stars (spread over the whole sky) will be observed at 2-min cadence. \tess\ observers 13 sectors per hemisphere, per year for at least ${\approx27}$ days; where the sectors overlap near the ecliptic poles, in the two Continuous Viewing Zones (CVZs), \tess\ observes for up to ${\approx350}$ days.  The CVZs are especially valuable places to search for exoplanets as the longer baseline enables the detection of smaller and/or longer-period planets---like the CBP presented here---and also overlaps with the James Webb Space Telescope CVZ \citep{Ricker2015}. Half way through its primary mission, \tess\ has already discovered a number of confirmed planets and identified more than a thousand planet candidates \citep[e.g.\ ][and references therein]{Huang2018,Vanderspek2019,Kostov2019}, vetted by the \tess\ 
Data Validation initiative \citep{Twicken2018, Li2019} (Guerrero et al., in prep.) 

\subsection{Discovery of the Host Eclipsing Binary\label{subsec:EB_discovery}}

TOI-1338 was identified as an eclipsing binary in 2009 as a part of the EBLM \citep[Eclipsing Binary Low Mass;][]{2013A&A...549A..18T} project, a survey constructed using the false-positives of the WASP survey for transiting hot-Jupiters  \citep{2006PASP..118.1407P, 2007MNRAS.380.1230C, 2011PhDT........22T}. As the observed eclipse depth is comparable to that of an inflated hot Jupiter, at the time it was not possible to distinguish between an eclipsing late-type M-dwarf and a transiting hot-Jupiter from the photometric signature alone. Follow-up observations with the CORALIE\footnote{CORALIE is a fibre-fed \'echelle spectrograph, mounted on the Swiss Euler 1.2 metre telescope at La Silla, Chile. It has a resolving power of $R=55,000$, and achieves long-term stability through a thermally-stabilised housing and nightly calibrations with respect to a Thorium-Argon reference spectrum \citep{2007A&A...468.1115L}.} high-resolution spectrograph obtained near quadrature revealed a semi-amplitude of 21.6 ${\rm km~s^{-1}}$, confirming that the target is a low-mass eclipsing binary \citep{2017A&A...608A.129T}. As part of the EBLM project to improve the M-dwarf mass-radius-temperature-luminosity relation, \tess\ 
observations at two-minute cadence were obtained for the target under a Cycle 1 Guest Observer program for Sectors 4-12 (G011278---PI O.~Turner).

\subsection{Detection of the Circumbinary planet\label{subsec:planet_discovery}}

Based on the two-minute cadence observations from \tess, the target was flagged as an eclipsing binary on the Planet Hunters TESS platform (Eisner et al. 2020)\footnote{\url{https://www.zooniverse.org/projects/nora-dot-eisner/planet-hunters-tess}} by user Pappa on subject 31326051. This is a citizen science project where, in addition to primarily flagging transit-like features, volunteers may flag targets as various phenomena including eclipsing binaries, variable stars, etc. Planet Hunters has had successful contribution to the field of CBPs through the independent discovery of Kepler-64 
\citep[also known as Planet Hunters-1,][]{Kostov2013,Schwamb2013}.

As part of our search for CBPs around eclipsing binaries from \tess, 
we have been performing a visual inspection of targets tagged as potential eclipsing binaries (with a hashtag ``eclipsingbinary") on the Planet Hunters \tess\ Talk. We note that TOI-1338 was also listed on exo.MAST\footnote{\url{https://exo.mast.stsci.edu}} 
as two Threshold Crossing Events with a period of $\sim 14.61$ days.

During the examination of the light curves on Planet Hunters \tess\ Talk, one of us (WC) noticed that in one of the partial light curves of TOI-1338 there was both a prominent primary eclipse and an additional feature. The latter was of similar depth to the secondary eclipse, but not at the expected time according to exo.MAST. To confirm that the feature is real and not a false positive caused by instrumental artifacts, we next extracted the 2-min cadenence light curve using the 
\lightkurve\ software package \citep{Lightkurve2018} 
and confirmed that there are indeed two genuine transit-like features in Sectors 6 and 10 that are not associated with the secondary eclipses, separated by ${\approx 95}$ days, and with different durations (${\sim0.3}$ and ${\sim0.6}$ days respectively). Further analysis of the 30-min cadence data, extracted with \eleanor\ \citep{feinstein2019}, revealed the presence of a third transit in Sector 3, ${\approx 93}$ days before the transit in Sector 6, with a duration of ${\sim0.4}$ days, further strengthening the CBP interpretation. Overall, the three transits exhibit the trademark ``smoking gun'' signatures of transiting CBPs \citep{Welsh2018} where (i) the transit durations vary depending on the orbital phase of the host EB such that transits across the primary star occurring near primary eclipses 
have shorter durations than transits occurring near secondary eclipses, and (ii) the 
transit times vary significantly from a linear ephemeris where specifically in this case the interval between the first and second transits ($\sim 93$ days) is significantly different than the interval between the second and third transits ($\sim 95$ days).

We note that it is highly unlikely for the three CBP transits to be a false positive scenario due, for example, by an unresolved eclipsing binary star. First, because the three transits have different durations, there would be only two plausible scenarios. One scenario, S1, involves two unresolved eclipsing binaries where one (hereafter EB1) would produce the first and third transits as primary and secondary eclipses, and the other (hereafter EB2) produces only the second transit as either a primary or a secondary eclipse. The other scenario, S2, requires an unresolved triple star system consisting of EB1 (producing the first and the third transits as a primary and secondary eclipse) and a long-period third star (hereafter EB3) producing the second transit as an eclipse across either the primary or the secondary star of EB1. For circular orbits, EB1 needs to have an orbital period of about 400 days (twice the time between the first and third transit) regardless of the scenario, EB2 for scenario S1 should have an orbital period greater than about 200 days (so that it would not produce a second eclipse during the TESS observations), and EB3 would need to have a dynamically-stable orbit around the ${\sim 400}$-day EB1. Second, because the three CBP transits have a depth of ${\sim0.2\%}$, such background EBs cannot be fainter than ${\rm T\sim18}$ mag (i.e.~ ${\rm \Delta T\sim7}$ mag difference compared to TOI-1338). We note that spectroscopy observations of the target do not show any signs of a second or third EB, albeit not as strong as 7 magnitudes difference. Below we explore this further using rough approximations to estimate the order of magnitude of the probability.

While the TESS Input Catalog indicates that there are 98 contaminating sources for TOI-1338, Gaia shows that there are only 6 sources within ${\rm \Delta T\sim7}$ mag\footnote{Gaia magnitude is similar to TESS magnitude,~i.e. $G = T + 0.43$ (Stassun et al. 2019)} inside the entire $13 \times 13$ TESS pixel array of the target, although none of them is inside the TESS aperture of the target. Assuming that these 6 sources are representative of the field of view, the density of sources within ${\Delta T\sim7}$ mag of the target is then ${\sim 0.036}$ sources/pixel, i.e. ${\sim 10^{-4}}$ sources/sq. arcsec. The contrast sensitivity of Gaia DR2 for ${\Delta G\sim7}$ mag (and thus ${\rm \Delta T\sim7}$ mag) is ${\sim 3}$ arcsec (Brandeker \& Cataldi 2019). Thus the probability of having one ${\Delta T\sim7}$ mag source unresolved by Gaia within 9 sq. arcsec of the target is ${\rm P_{unresolved}\sim 10^{-3}}$. 

Using the results of Raghavan et al. (2010), we estimate that the probability of EB1 having an orbital period of $400$ days is ${\rm P_{\rm EB1, per}\sim20\%}$; the probability of EB2 having an orbital period of $200$ days is ${\rm P_{\rm EB2, per}\sim85\%}$. Using the results of Tokovinin (2014), we estimate the probability of a triple star to be ${\rm P_{triple}\sim 10\%}$. The probability that EB1, EB2, and EB3 are eclipsing is roughly ${\rm (R_1 + R_2)/a}$. Assuming similar stars (because of equal depth eclipses) of solar-radius and mass (because larger stars are more rare and smaller stars would be too faint to produce the required contamination) for both scenarios, the corresponding probabilities are ${\rm P_{\rm EB1, ecl}\sim (2\times 0.0046)/1.34\sim0.006}$ and ${\rm P_{\rm EB2, ecl}\sim (2\times 0.0046)/0.84\sim0.011}$. For scenario S2, we used Rebound's IAS15 integrator (Rein \& Spiegel 2015) to test that the orbital period of EB3 would need to be greater than ${\sim2000}$ days to be dynamically stable. Thus the probability that EB3 is eclipsing EB1 is ${\rm P_{\rm triple, ecl}\sim (2\times 0.0046)/4.48\sim0.002}$. 

We also note that the orbital phases of EB1, EB2, and EB3 would need to be such that in ${\sim 325}$ days (duration of TESS observations) EB1 would produce one primary and one secondary eclipse, and EB2 and EB3 would produce a single eclipse---within a window of a few days for its duration to agree with the CBP model. This introduces additional constraints. Namely, for orbital periods of 400-days, 200-days, and 2000-days for EB1, EB2, and EB3 respective, and assuming said window is ${\sim 2}$ days, the corresponding probabilities are ${\rm P_{EB1,phase}\sim80\%}$ (i.e.$\sim$325/400), ${\rm P_{EB2,phase} \sim1\%}$ (i.e.$\sim$2/200), and ${\rm P_{EB3,phase}\sim0.1\%}$ (i.e.$\sim$2/2000). 

Putting all this together, the combined probability for scenarios S1 and S2 is as follows:

\begin{equation}
        {\rm P_{\rm S1} = P_{unresolved, EB1}\times P_{unresolved, EB2}\times P_{\rm EB1, per}\times P_{\rm EB2, per}\times P_{\rm EB1, ecl}\times P_{\rm EB2, ecl}\times P_{EB1,phase}\times P_{EB2,phase} = }
\end{equation}

\begin{equation}
        {\rm  = 10^{-3}\times10^{-3}\times0.2\times0.85\times0.006\times0.011\times0.8\times0.01 \sim 10^{-13}}
\end{equation}

\begin{equation}
        {\rm P_{\rm S2} = P_{unresolved}\times P_{triple}\times P_{\rm EB1, per}\times P_{\rm EB1, ecl}\times P_{\rm EB3, ecl}\times P_{EB1,phase}\times P_{EB3,phase} = }
\end{equation}

\begin{equation}
        {\rm  = 10^{-3}\times0.1\times0.2\times0.006\times0.002\times0.8\times0.001 \sim 2\times10^{-13}}
\end{equation}

To convert these numbers to a false positive probability (FPP), we compare them to the probability that TOI-1338 b is a circumbinary planet. Specifically, from the Kepler dataset the probability that any given star has a transiting CBP is ${\sim 10/200,000 = 5\times10^{-5}}$. Thus the CBP hypothesis is ${\sim 5\times10^{-5}/10^{-13} = 5\times10^{-8}}$ times more likely than the false positive hypothesis. The FPP is therefore about ${10^{-8}}$.

These false positive scenarios can be further argued against based on the durations and depths of the three transits. Specifically, as the depths of the first and third transits are similar, then the two stars of EB1 should have comparable sizes. As discussed above, assuming Sun-like stars the semi-major axis of EB1 would be $\sim1.34$ AU and the corresponding duration of the primary eclipse of EB1 (for circular orbit, $R_1 = R2 = R_\odot$, and impact parameter b = 0) would be $\sim0.9$ days, i.e.~nearly twice the duration of the observed transits, thus further ruling out this specific scenario. While an eccentric orbit of EB1 might alleviate this tension to some degree, this would require special orbital elements -- in addition to the requirement imposed by the special orbital phase as discussed above. Thus overall we consider false positive scenarios S1 and S2 to be highly unlikely. 



\subsection{\tess\ Lightcurve}

\begin{figure}
    \centering
    \includegraphics[width=0.9\textwidth]{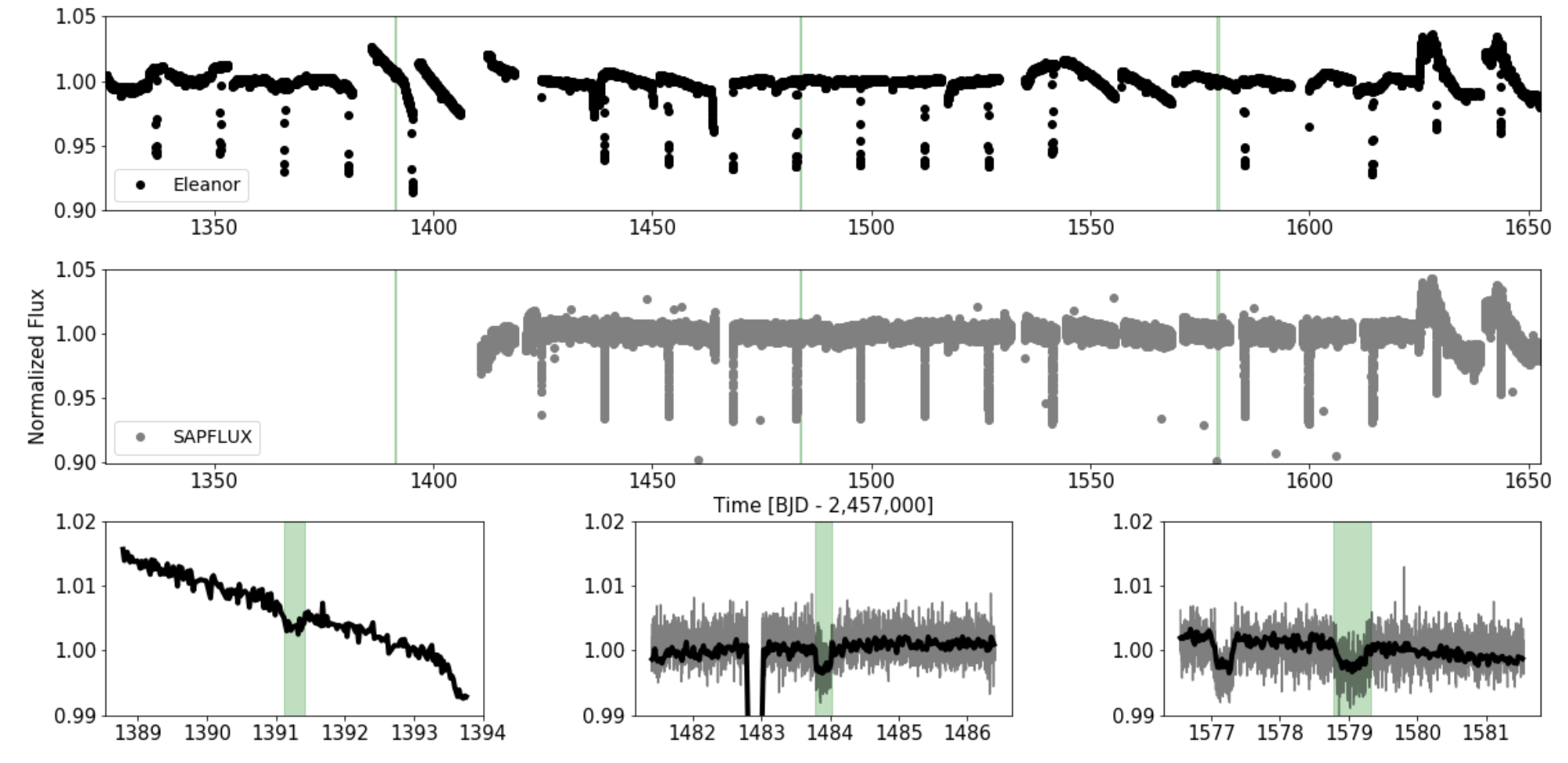}
    \caption{Upper panel: Sectors 1-12 of 30-min cadence \tess\ \eleanor\ PSF-extracted lightcurve (black). Some of the eclipses were missed due to data gaps. The CBP transits are highlighted in green; Middle panel: Sectors 4-12 of 2-min cadence SAPFLUX lightcurve (gray); Lower panels: 5-day sections of the lightcurve centered on the three CBP transits (highlighted in green). The middle and right panel also show primary and secondary eclipses near days 1483 and 1577, respectively.\label{fig:LC_fig_full}}
\end{figure}

\tess\ telemeters data in two modes: postage stamps, i.e. small regions, around roughly 20,000 stars at 2-minute cadence every sector\footnote{For sectors 1-3 slightly less than 16,000 targets received 2-min cadence observations. The number of 2-min cadence targets was increased to 20,000 after predicted compressibility was demonstrated in flight and the SPOC demonstrated in a ground segment test that 20,000 targets could be handled in the pipeline \citep{Jenkins2016}.} as well as the Full-Frame Images, which contain about a million stars each (brighter than T = 15 mag), at 30-minute cadence. TOI-1338 was observed by \tess\ in 30-min cadence in Sectors 1-12, and in 2-min cadence in Sectors 4-12.  

The \tess\  
light curve of TOI-1338 is shown in Figure \ref{fig:LC_fig_full}, where the 30-min cadence
data extracted with \eleanor\ using aperture
photometry is shown for Sectors 1-12 and 2-min cadence SAPFLUX measurements
from the standard processing provided by the \tess\ mission is shown for Sectors 4-12. The nominal units of the observation times are days
from BJD 2,457,000. The \eleanor\ software package performs background subtraction, aperture photometry, and detrending for a given source on the FFIs. It also provides the opportunity to use a custom aperture and if desired can use models of the point spread function (PSF) to extract the light curves. 

Each full-frame image delivered by the \tess\ project is barycentric corrected. However, there is only a single correction applied to each CCD, each of which covers more than 100 square degrees of the sky. As a result, the barycentric-corrected times in the raw FFI data can be discrepant by up to a minute. The \eleanor\ software corrects for this potential offset, removing the barycentric correction and applying a more accurate value given the actual position of the target in question. We verified the \eleanor\ timestamps were accurate, comparing them to the midpoint of the 15 2-minute images that make up a single FFI. We find the two are consistent at the 2-second level, sufficiently precise for the photodynamical modeling we employ in Section \ref{sec:ELC}.

In addition to the prominent stellar eclipses the light curve of TOI-1338 contains several small, transit-like events that required further scrutiny. Specifically, we noticed four events near days 1390, 1391, 1403, and 1404 (Sector 3), one event near day 1484 (Sector 6), and an event near day 1579 (Sector 10). Using the PSF fitting built into \eleanor\, we showed that the events near days 1391, 1484, and 1579 are astrophysical in origin---namely, the three transits of the CBP---while the remaining events are artifacts caused by
pointing ``jitter'' (mainly before reaction wheel ``momentum dumps'').

\begin{figure}
    \centering
    \includegraphics[width=0.75\textwidth]{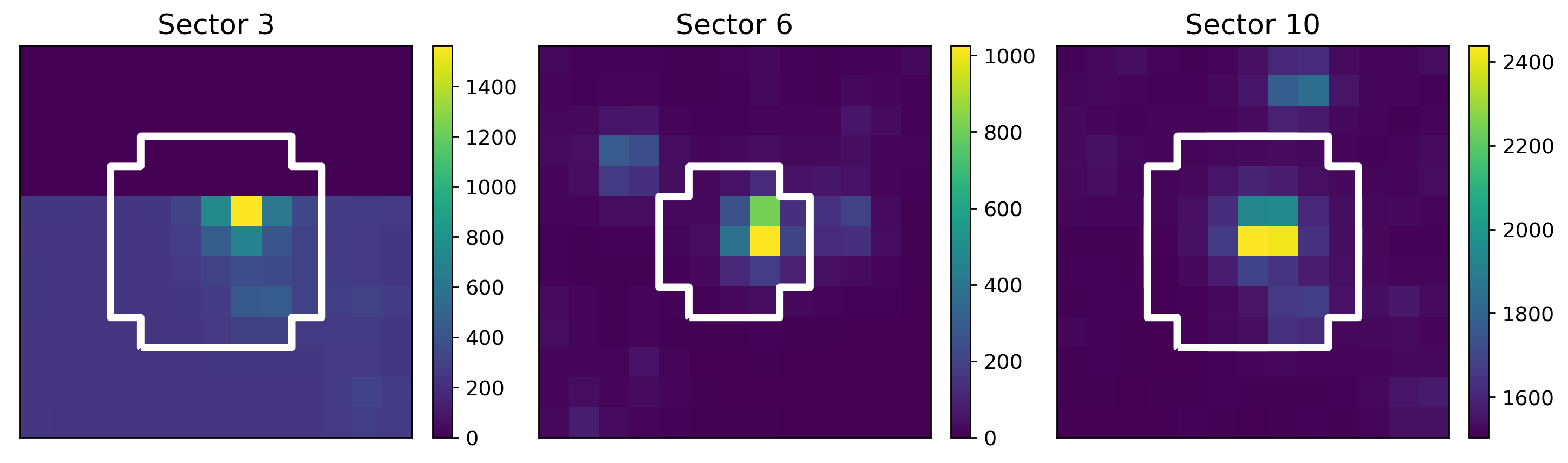}
    \caption{Nominal \eleanor\ apertures for Sectors 3, 6, and 10. 
    The target was on the edge of the detector in Sector 3, 
    as seen in the left panel.  Hence in this case a custom aperture is needed.\label{fig:eleanor_apertures}}
\end{figure}

The 30-min cadence light curve from Sector 3 was particularly difficult to extract
because the
target fell near an edge of the detector as shown in Figure
\ref{fig:eleanor_apertures} (the target was well away from
the detector edges in the remaining Sectors).
Additionally, the light curve extraction is especially prone to systematic
errors as occasional spacecraft pointing
jitter during Sector 3 moved some of the
light from the target off of the detector, thereby
producing transit-like events in the observed flux.
As discussed below, the PSF of the target changes during some of the
events listed above, which helps us rule out an astrophysical
origin.

\begin{figure}
    \centering
    \includegraphics[width=0.75\textwidth]{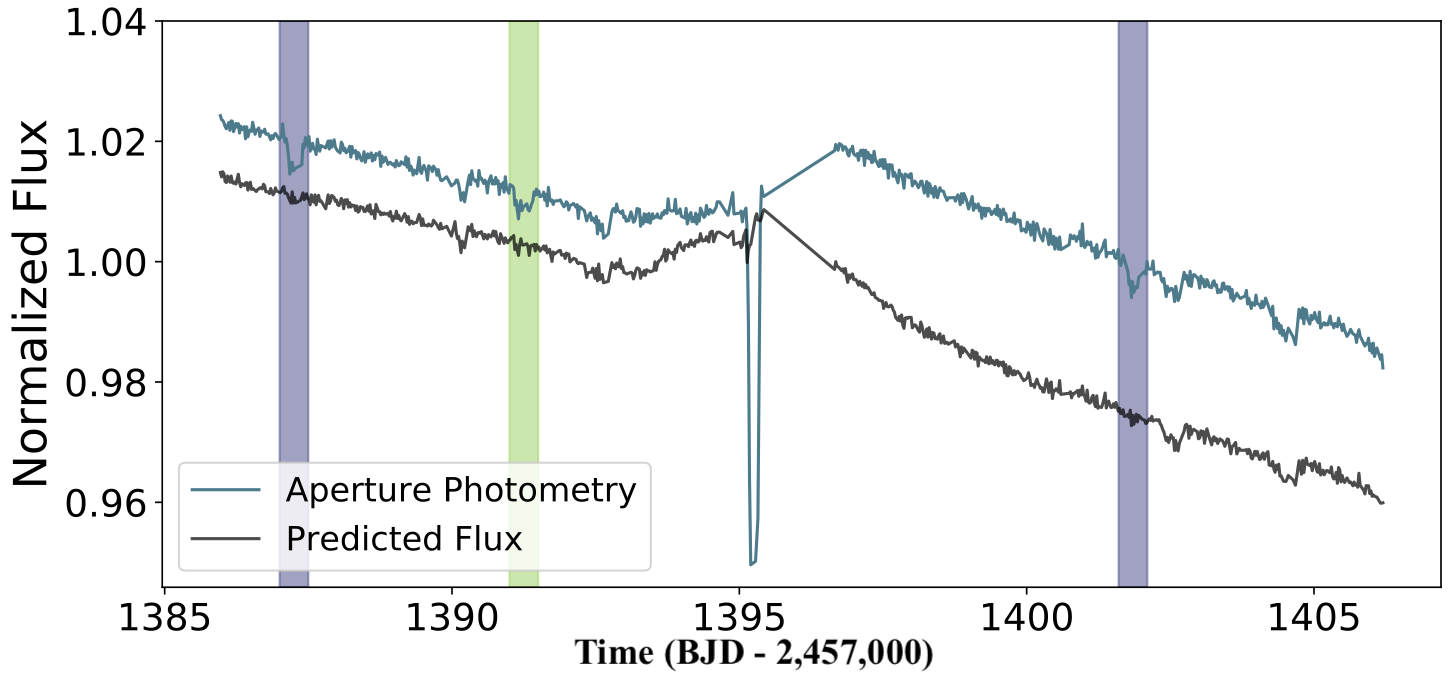}
    \caption{Aperture photometry for Sector 3 data (upper
    curve) with the flux model predicted from changes in the shape of the PSF at each cadence inferred by \eleanor\ PSF modeling (lower curve). We expect that spurious events caused by changes in the PSF to appear in both curves, whereas astrophysical events, which occur independently of the instrumental PSF, will only appear in the upper curve.  Thus, the apparent events near days 1390, 1403 and 1404 are caused by changes in the PSF shape which occur as parts of the star fall off and on the detector owing to pointing jitter. The purple- and green-shaded regions correspond to the locations of secondary eclipses and a transit of the planet, respectively.  These events are apparent only in the upper curve, which is a strong indication of their validity. Finally, we note the deep feature near day 1395 is a primary eclipse.}\label{fig:sect3_fake}
\end{figure}

\begin{figure}
    \centering
    \includegraphics[width=0.99\textwidth]{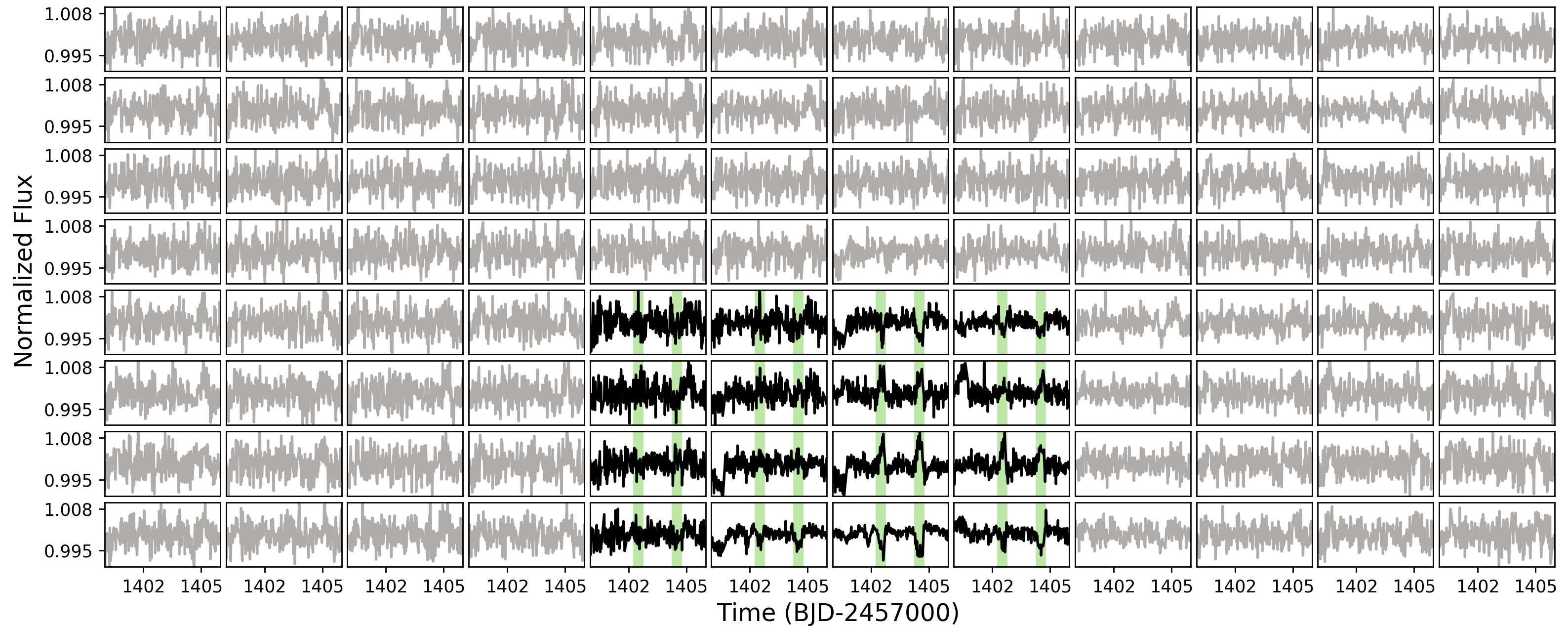}
    \caption{A section of the \tess\ \eleanor\ per-pixel light curve for Sector 3 near days 1402 and 1405. The panels with black light curves represent the core pixels used for light curve extraction. The two events highlighted in green appear as transit-like in some of the core pixels and anti-transit-like in others. This demonstrates that these two events are data artifacts (see Fig.\ \ref{fig:sect3_fake}).\label{fig:sect3_pbp}}
\end{figure}

The PSF can be modeled in \eleanor\ using either a two-dimensional
Gaussian or a Moffat profile, and in this case
both models perform equally well.
Briefly, the analysis proceeded as follows. At each cadence, we fit parameters that describe the shape of the stellar PSF, assuming all stars in a $13\times13$ pixel region share the same PSF. We then optimized the flux of each star, the shared PSF parameters, and a single background level across this region. 
Based on this analysis, we found that during the events near days 1390, 1403 and 1404 in Sector 3, the shape of the PSF changed, affecting how much of the star falls in the optimal aperture and therefore how much flux is observed. It is likely that the pointing became ``looser'' during the times of these events, and the PSF-fitting package interpreted the resulting images as an increase in the size of the PSF. The final pointing algorithm was implemented on board the spacecraft after Sector 4, and TESS experienced sporadic pointing errors more frequently and of higher amplitude prior to that. Owing to an unfortunate set of coincidences, the spurious transit events in Sector 3 happen to
have similar depths and durations as the real CBP transits and secondary eclipses. Given the PSF change,
we then built a linear model from the out-of-eclipse data that predicts the flux of the star at every cadence from the PSF parameters alone. 
Figure~\ref{fig:sect3_fake} shows part of the Sector 3 light curve
from \eleanor\ using aperture photometry (top curve) and
the model light curve predicted from changes in the PSF
(bottom curve).
We found that the PSF-model predicts transit-like features 
near days 1390, 1402, and 1405. However, this model  
does {\em not} predict transit-like events at the time of a
secondary eclipse (days 1387 and 1402) or near day 1391 (CBP transit). This is further demonstrated in Figure~\ref{fig:sect3_pbp}, showing a section of the per-pixel light curve for the two events near days 1402 and 1405---the events are anti-correlated among some of the core pixels such that they appear as transit-like in some pixels and anti-transit-like in others, indicating data artifacts. We thus confirmed the reality of the day 1391 transit-like event (and also the secondary eclipses near days 1387 and 1402) and conclude that the apparent transit-like events near days 1390, 1403 and 1404 are instrumental artifacts. A similar analysis confirmed the reality of the two CBP transits in Sectors 6 and 10.
%
%
We found the long- and short-cadence light curves to have different eclipse depths in some sectors because of different levels of background subtraction. In many cases, the short cadence data overestimated the background, causing many ``sky'' pixels near the star to record negative flux values and the eclipses to appear artificially deep. We re-fit a background model for the short cadence data using the full-frame images, interpolating to the times of each short cadence exposure, leading to consistent eclipse depths between the two datasets. We also tested a variety of different apertures, finding the choice of aperture did not make a significant difference on the ultimate photometry, and in most sectors we used the pipeline default aperture.

We achieved the best photometric precision for \target\ 
using the PSF-based photometry built into \eleanor.  Our final
adopted light curve is a combination of 30-min cadence data extracted
from the FFIs 
for Sectors 1, 2, and 3, and 2-min cadence data extracted from the 
target pixels in the ``postage stamps'' for this particular target for the remaining sectors.
This final light curve was
detrended and normalized in the manner described in
\citet{Orosz2019}. As part of this iterative detrending process, we measure the durations of all eclipse and transit events. Most of the out-of-eclipse portions of the light curve are trimmed, and we keep only the out-of-eclipse regions that are within 0.25 days or $1.25\tau$ of each event (which ever is greater), where
$\tau$ is the duration of the event.  The last primary eclipse in Sector 12 had unusually large residuals (for unknown reasons), and consequently we excluded it from further analysis. 

We measured the times of the eclipses and transits by fitting
a simple model to trimmed and normalized light curves for each event
using the Eclipsing Light Curve (ELC) code of Orosz \& Hauschildt (2000). This model has nine free parameters: the period $P$, the conjunction time 
$T_{\rm conj}$, the inclination $i$, the primary radius $R_1$, the
ratio of the radii $R_1/R_2$, two limb darkening parameters for
the quadratic limb darkening law $q_{1,1}$, $q_{2,1}$, and two eccentricity parameters $\sqrt{e}\cos\omega$ and $\sqrt{e}\sin\omega$. The goal was to find a smooth and symmetic curve that best fit each segment, so no attempt was made to optimize more than one segment at a time.  For each segment, we found the best-fitting model using the Differential Evolution Monte Carlo Markov Chain (DE-MCMC) algorithm of Ter  Braak (2006) with 80 chains that was run for 5000 generations.  Using the best-fitting model, the individual uncertainties on each point were scaled in such a way to get $\chi^2_{\rm best}=N$, where $N$ is the number of points.  The scale factors for the first five primary eclipses (observed in long cadence) were 0.347, 0.353, 0.626, 0.690, and 0.770, respectively. For primary eclipses observed in long cadence, the scale factors ranged from 1.042 to 1.376 with a median of 1.121.  After the
uncertainties on the data in each segment were scaled, the 
DE-MCMC code was run using 80 chains for 40,000 generations.  Posterior samples were selected starting at generation 400, and sampling every 400th generation thereafter.  The median of each posterior sample was adopted as the eclipse time, and the rms of the sample was taken to be the $1\sigma$ uncertainty.    The measured primary and secondary cycle numbers and times
are given in Table \ref{eclipsetimes1}.  The cycle numbers for the
secondary eclipses are given as fractional values of the form NN.45345 (the mean phase of the secondary eclipses is not 0.5 owing to the orbital eccentricity).

To make the final light curve that was modeled using the photodynamical
model described in Section \ref{sec:ELC} below, we simply combined the individual segments with scaled uncertainties from all of the eclipse and transit events.  Since each segment has had its uncertainties scaled individually, no one event should be given undue weight owing to underestimated uncertainties for the photometric measurements.  The scale factors for the segments in long cadence were smaller than one, which suggests the photometric errors were overestimated.
On the other hand, the scale factors for the event observed in short cadence were all larger than one, which suggests the photometric errors were underestimated.

\section{Complementary Observations\label{sec:complementary}}

\subsection{Radial Velocity Characterization}

After its initial classification as an eclipsing binary, EBLM J0608-59 was observed with the CORALIE spectrograph to constrain the component masses. Between December 2009 and April 2012, 19 radial velocity observations were performed to map out the Keplerian orbit of the binary
\citep{2017A&A...608A.129T}. Exposures were typically 600 s long, yielding a median precision of 34 m s$^{-1}$. Owing to the small mass ratio of the binary, and indeed of the entire EBLM sample by construction, the system appears as a single-line spectroscopic binary. We expect a $V$ magnitude difference of 9.8 between the primary  
and the secondary, given that $M_{1}\sim 1.1\,M_{\odot}$, $M_{2}\sim 0.3\,M_{\odot}$,
respectively
\citep[][]{2017A&A...608A.129T}. As a consequence of the large flux ratio between the primary and secondary, 
the spectral lines of the secondary are not noticeable in the observations, thereby allowing high-accuracy measurements of the primary's radial velocity.
Likewise, the secondary star is too faint for \tess\
to be able to detect transits of the CBP across it. 

\subsubsection{The BEBOP radial velocity search for circumbinary planets}

In late 2013 the BEBOP (Binaries Escorted By Orbiting Planets) program was created as a radial velocity survey for the detection of circumbinary planets. The BEBOP sample is exclusively constructed from eclipsing single-line spectroscopic binaries to avoid contamination effects that severely hinder planet detection in double-line spectroscopic binaries \citep{2009ApJ...704..513K}. An initial target list of roughly 50 binaries was created from the larger EBLM sample. Selection criteria include the obtainable radial-velocity precision and the lack of stellar activity. EBLM J0608-59 was included in this initial BEBOP selection. Between November 2014 and August 2015 we acquired 17 additional CORALIE measurements with longer exposures of 1800 s to increase the  precision of the radial velocity measurements (the median uncertainty was 25 ms$^{-1}$). This second set of observations coincided with a change to a new octagonal fibre. The new fibre provided greater long-term stability compared to the original circular one, but at the cost of $\sim 10\%$ of the incoming flux. Such a fibre change may also induce a small velocity offset \citep{2017MNRAS.467.1714T}. Consequently, when modelling the data we treat the CORALIE data sets as if they were acquired from two different instruments, with a free parameter for the offset.

The combined radial velocities from both the EBLM and BEBOP surveys are presented in \citep{Martin2019}. No circumbinary planet was detected in the radial velocity of this system within the sensitivity of the observations, which were adequately fit by a single Keplerian orbit ($\chi^2_{\nu} \sim 1$). This ruled out (at 95\% confidence) the presence in the system of a CBP more massive than $\sim 0.7M_{\rm Jup}$ and with orbital period of roughly six times shorter than that of the binary.

Starting in April 2018 the BEBOP survey was extended to the HARPS high-resolution spectrograph on the ESO 3.6-m telescope \citep[Prog.ID 1101.C-0721, PI A. Triaud;][]{2002Msngr.110....9P}, also at La Silla, Chile\footnote{BEBOP also surveys the northern skies, using SOPHIE, at the Observatoire de Haute-Provence, Prog.ID 19A.PNP.SANT, PI A. Santerne. The northern EBLM sample will also be included in the \tess\
2 minute cadence under proposal G022253, PI D. Martin.}. Compared to CORALIE, HARPS benefits from a larger telescope
aperture, a higher resolving power of $R=115,000$, and greater radial-velocity stability by being both thermally-stabilized {\it and} operated under vacuum. Seven HARPS spectra of the target have been acquired to date. They were reduced with the HARPS pipeline, which has been shown to achieve remarkable precision and accuracy \citep[e.g.][]{2009A&A...493..639M,2014ApJ...792L..31L}. The radial-velocities were computed by using a binary mask corresponding to a G2 spectral-type template \citep{1996A&AS..119..373B}. We achieve a median radial-velocity precision of 5.9 m$^{-1}$ in our measurements. A fit to the complete set of radial-velocities (CORALIE and HARPS) produces a fit of reduced $\chi^2_{\nu} \sim 1$ mostly because the number of HARPS measurements is close to the number of free parameters for the binary orbit. The model adjusts to the HARPS measurements first, because they have the greatest weights. CORALIE's precision is not sufficient to detect the additional planetary signal. The HARPS measurements indicate that this system has very little activity, making it optimal for radial-velocity measurements. The BEBOP survey is ongoing and more spectra from both HARPS and ESPRESSO will be obtained for TOI-1338/EBLM J0608-59, and published in a subsequent paper.

\begin{figure}%
\centering
\includegraphics[height=1.9in]{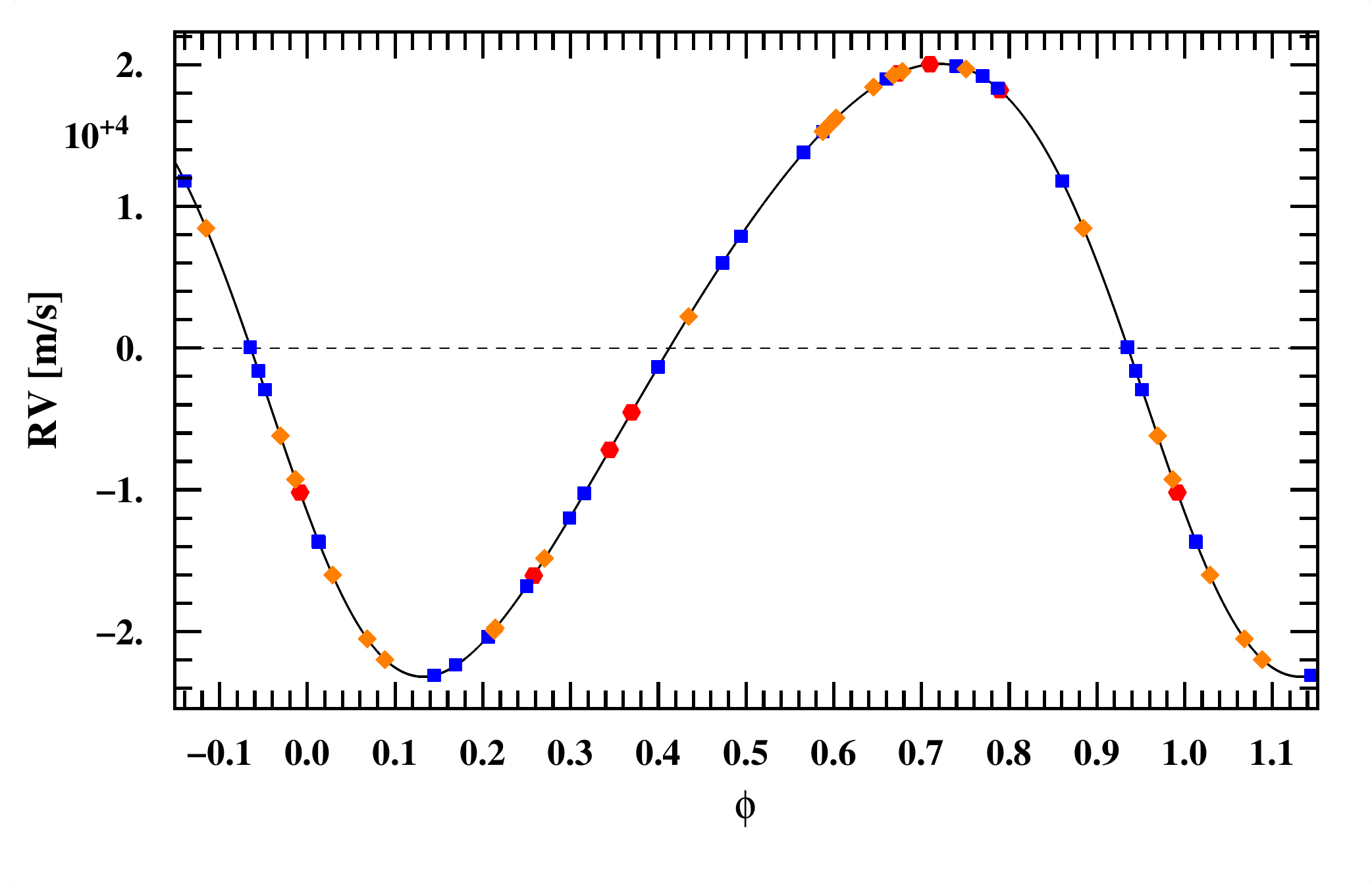}%
\qquad
\includegraphics[height=1.9in]{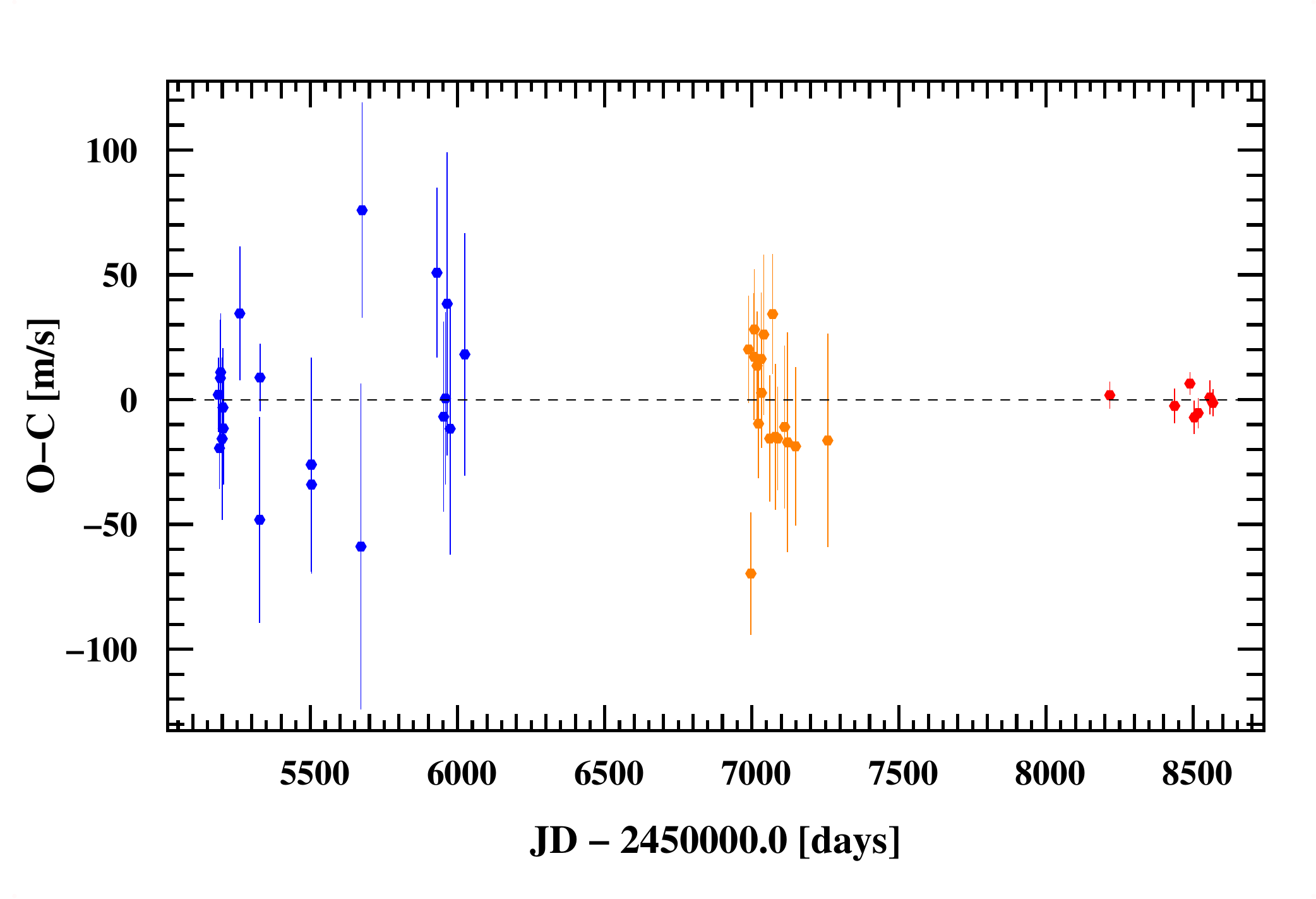}%
\caption{Left: radial velocities for TOI-1338, phase-folded on the binary orbital period of 14.61 days, taken with the CORALIE instrument (blue and orange) and HARPS (red). Right: residuals to the single-Keplerian radial velocity fit over time. The change in CORALIE residuals between blue and orange measurements coincide with a fibre change (see text for details).\label{fig:RVs}}
\end{figure}

\subsection{Spectroscopic Characterization}

The seven extracted HARPS spectra were co-added onto a common wavelength axis reaching a signal-to-noise ratio of approximately 73. The resulting spectrum was analysed with the spectral analysis package \textsc{ispec} \citep{2014A&A...569A.111B}. We used the synthesis method to fit individual spectral lines of the co-added spectra. The radiative transfer code SPECTRUM \citep{1994AJ....107..742G} was used to generate model spectra with MARCS model atmospheres \citep{2008A&A...486..951G}, version 5 of the GES (GAIA ESO survey) atomic line list provided within \textsc{ispec} and solar abundances from \citet{2009ARA&A..47..481A}. Macroturbulence is estimated using equation (5.10) from \citet{2015PhDT........16D} and microturbulence was accounted for at the synthesis stage using equation (3.1) from the same source. The H$\alpha$, Na\,I\,D and Mg I b lines were used to infer the effective temperature $T_{\rm eff}$ and gravity $\log g$ while Fe\,I and Fe\,II lines were used to determine the metallicity [Fe/H] and the projected rotational velocity $\rm v sin i_\star$. Trial synthetic model spectra were fit until an acceptable match to the data was found. Uncertainties were estimated by varying individual parameters until the model spectrum was no longer well-matched to the spectra of TOI-1338. For the primary star, we find an effective temperature of $T_{\rm eff,1}  = 6050 \pm 80$~K, a metallicity of [Fe/H]$_1 = 0.01 \pm 0.05$,  a gravity of $\log g_1 = 4.0 \pm 0.08$ dex, and a projected rotational velocity of  $v \sin i_{\star, 1} = 3.6 \pm 0.6$ km s$^{-1}$. These measurements are summarized in Table \ref{tab:pre_ELC}.

As a check on the spectroscopic temperature of the primary, we first gathered brightness measurements from the literature in the Johnson, Tycho-2, 2MASS, and Sloan systems, constructed 10 non-independent color indices, and corrected each for reddening using the extinction law of \cite{Cardelli1989} with a value of $E(B-V) = 0.020 \pm 0.010$ derived from the extinction map of \cite{Schlafly2011} and the {\it Gaia} \citep{Gaia2016,Gaia2018} distance
of $405\pm 3$ pc.
These colors are unaffected by the secondary star because it is so faint. We then used color-temperature calibrations from \cite{Casagrande2010} and \cite{Huang2015} to infer a mean photometric temperature of $5990 \pm 110$~K, in good agreement with the spectroscopic value.

We estimated the radius of the primary star in three different ways. One method used the procedure outlined by \cite{Stassun2019} for the preparation of the TIC-8 catalog, involving the {\it Gaia\/} parallax and $G$ magnitude, an extinction correction, the spectroscopic temperature, and the $G$-band bolometric correction. Another method we used was based on a fit to the spectral energy distribution (SED) performed with EXOFASTv2 \citep{Eastman2019} and the MIST bolometric correction tables\footnote{\url{http://waps.cfa.harvard.edu/MIST/model_grids.html}}, using brightness measurements in the Gaia and WISE systems in addition to those mentioned earlier.  Suitable priors were placed on the distance, extinction, temperature, metallicity, and $\log g_1$. A third method we used was also based on an SED fit but used instead the NextGen model atmospheres of \cite{Allard2012}. The three procedures gave
very similar results. We adopt the value $R_1 = 1.345 \pm 0.046~R_{\sun}$ in the following to use as a prior for the photometric-dynamical modeling described below.

\subsubsection{Image Analysis}

\begin{figure}
    \centering
    \includegraphics{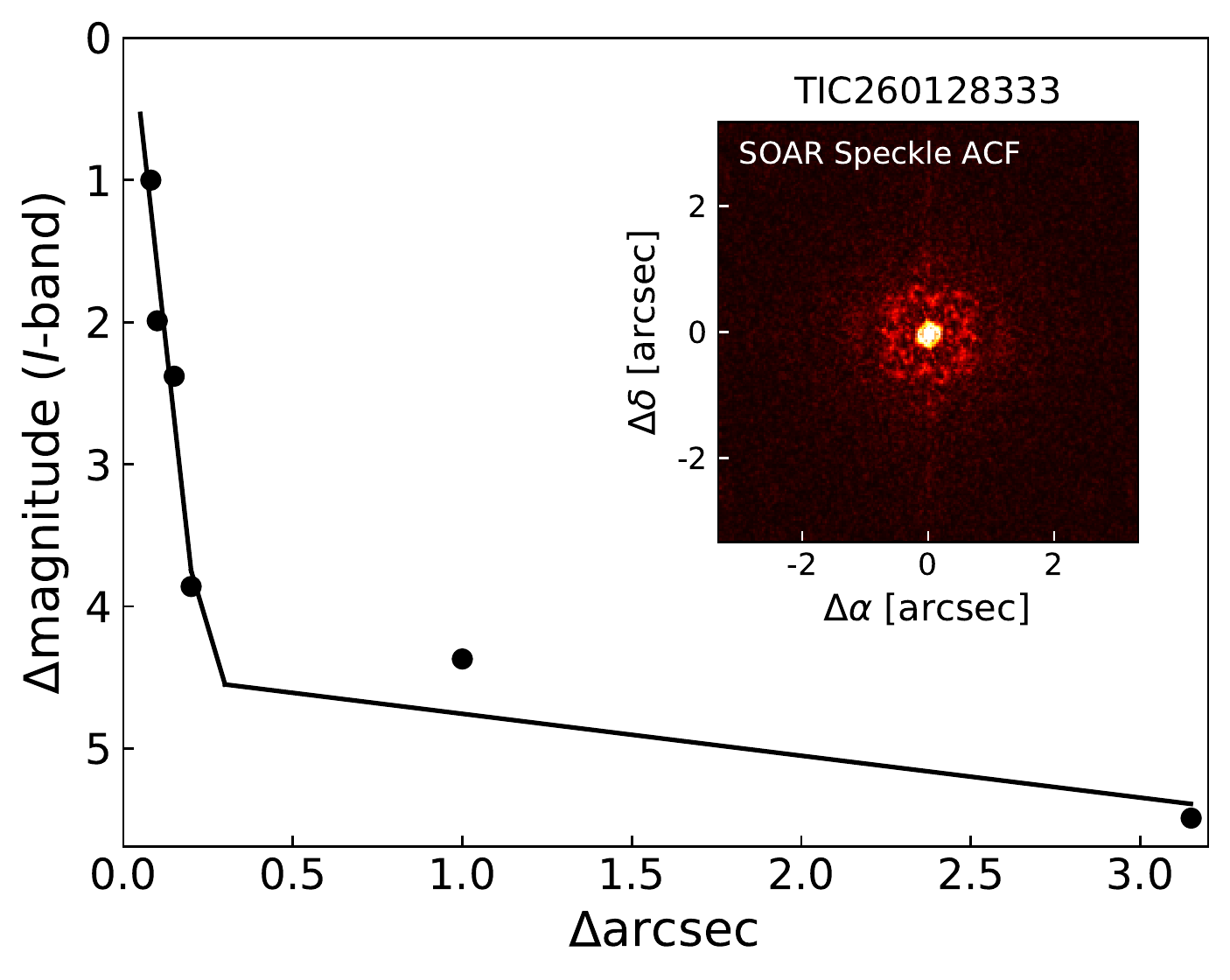}
    \caption{The  $5\sigma$ detection limits of the SOAR speckle imaging of TOI-1338; the inset is the speckle auto-correlation function. No nearby stars were detected within 3\arcsec\ of the target.\label{fig:SOAR}}
\end{figure}

The relatively large sizes of the \tess\ 
pixels, approximately 21\arcsec\ on a side, leave the target
susceptible to photometric contamination from nearby stars, including additional wide stellar companions. The nearest object (TIC 260128336) to TOI-1338 after proper motion correction is separated by 53.6\arcsec. The SPOC Data Validation centroid offsets for the Sectors 1-12 multi-sector run indicated that the primary and secondary eclipses for TOI-1338 originate from the target itself \citep{Jenkins2016}; complementary analysis with the photocenter module of {\textsf DAVE} (Kostov et al. 2019) confirmed this.  We also searched for nearby sources with SOAR speckle imaging \citep[see][for details of the instrumentation]{Tokovinin2018,Tokovinin2019} 
on 17 March 2019 UT, observing in a similar visible bandpass as \tess. We detected no nearby sources within 3\arcsec\ of TOI-1338. The detection sensitivity and speckle auto-correlation function of the SOAR observations are plotted in Figure \ref{fig:SOAR}.

\section{Photometric-Dynamical Analysis of the System\label{sec:ELC}}

Due to the rich dynamical interactions between the two stars and a planet in a CBP system, the deviations of the planet's orbit from a strictly periodic one are much more pronounced compared to a single-star system. The secular evolution in a CBP system can occur on a timescale as short as a decade instead of thousands of years (like in the Solar System).  Thus measurable changes in quantities such as the inclination of the planet's orbit can be observed with missions like {\em Kepler} and \tess. Given the relatively rapid secular evolution, the orbits in a CBP system are not simple Keplerians. A complete description of a CBP system relies on a large number of parameters---e.g.\ masses, radii, and orbital parameters for the two stars and the planet(s), radiative parameters for the two stars, etc.  Stellar eclipses, both primary and secondary, allow for precise measurements of times of conjunction.  In addition, transits of the CBP across the primary and/or the secondary star can provide precise position measurements of both the stars and the planet at times other than the times stellar conjunction.  Thus, the dynamical complexity of  the  system---while computationally challenging---enables precise measurements of the system's  parameters. For example, the stellar masses and radii of the two stars in the  Kepler-16 CBP system have been measured to sub-percent precision \citep{Doyle2011}.

\subsection{ELC Modeling} 

To obtain a complete solution for the TOI-1338 system, we carried out a photometric-dynamical analysis with the ELC code \citep{Orosz2000}, utilizing the photometry from \tess\ and the precise radial velocity measurements from CORALIE and HARPS. For this task, the ELC code combines N-body simulations, modified to include tidal interaction and general relativistic effects, with a photometric model for the stellar eclipses and planetary transits, to reproduce a light curve of the system. These modifications to ELC to allow for modelling stellar triple and higher-order systems, and CBP systems have been described in \citet{Welsh2015} and \citet{Orosz2019}. This code has been used 
extensively for the analysis and confirmation of {\it Kepler} CBPs; for the sake of completeness we outline it below.

Briefly, given instantaneous orbital parameters (e.g.\ the orbital period, the eccentricity, the inclination, etc.) at some reference epoch and the masses of each body, ELC solves the Newtonian equations of motion using a symplectic integrator, in this case a 12th order Gaussian Runge Kutta integrator 
\citet{Hairer2002}. When necessary, the Newtonian equations of motion can be modified to account for general relativistic precession and tidal effects \citep[see][]{Mardling2002}. 
The solution of the dynamical equations
enables the positions of all the bodies to be specified at any given time.  Then, given the positions of the bodies on the plane of the sky, their radii, and their radiative properties, the observed flux is computed for any number of overlapping bodies using the algorithm discussed in \citet{Short2018}.  Likewise, the solution to the dynamical equations gives the radial velocity of each body at any given time. It is also possible to include other observable quantities that do not depend on time (for example the surface gravity of the primary star) in the fitting process.

For TOI-1338, we initially had the following 25 free parameters. The binary orbit is specified by the time of conjunction $T_{\rm conj, bin}$, 
the period $P_{\rm bin}$, the eccentricity parameters 
$\sqrt{e_{\rm bin}}\cos\omega_{\rm bin}$\footnote{Combinations and/or ratios of individual parameters are often more convenient to use.} and
$\sqrt{e_{\rm bin}}\sin\omega_{\rm bin}$ 
(where $e_{\rm bin}$ is the eccentricity and 
$\omega_{\rm bin}$ is the argument of periastron), and the 
inclination $i_{\rm bin}$. We fix the 
initial nodal angle of the binary orbit to
$\Omega_{\rm bin}=0$. The stellar masses are specified by the primary mass $M_1$ (in units of $M_{\odot}$) 
and the mass ratio $Q=M_2/M_1$, the stellar radii---by the primary radius $R_1$ and the radius ratio $R_1/R_2$, and the stellar temperatures---by $T_{\rm eff, 1}$ and
the temperature ratio $T_{\rm eff, 2}/T_{\rm eff, 1}$. We assume
a quadratic limb darkening law with the \citet{Kipping2013}
``triangular'' resampling, for a total of four additional parameters.
Finally, to account for tidal precession we use two additional free parameters---the apsidal constants
$k_{2,1}$ and $k_{2,2}$ for the primary and secondary, respectively.
For the planet's orbit, we fit for the sum and differences of
the period and time of barycentric conjunction, i.e.
$\zeta_p=(P_{\rm pl}+T_{\rm conj, pl}$),
$\zeta_m=(P_{\rm pl}-T_{\rm conj, pl}$), the eccentricity 
parameters 
$\sqrt{e_{\rm pl}}\cos\omega_{\rm pl}$,
$\sqrt{e_{\rm pl}}\sin\omega_{\rm pl}$, the inclination
$i_{\rm pl}$, and the nodal angle $\Omega_{\rm pl}$.
The planet's mass is specified by $M_3$, where the units are
$M_{\oplus}$. Finally, the radius of the planet is specified by 
the ratio $R_1/R_{\rm pl}$. 


The observables for the TOI-1338 system are the trimmed and
normalized light curve from \tess, and 
the radial velocity measurements from CORALIE and HARPS.
The radial velocities span three distinct data sets where an offset velocity is found for each, namely the ``early'' CORALIE data before the fibre change (roughly days 200 to 1000 in units of BJD - 2,455,000), 
the ``late'' CORALIE data post fibre change (roughly days 1900 to 2250), and HARPS (roughly days 3200 to 3550).  
Another set of observables is the spectroscopically-measured gravity for the primary star ($\log g_1=4.0 \pm 0.08$ dex in cgs units), and the
radius of the primary star ($R_1=1.345\pm 0.046\,R_{\odot}$) as determined from multi-color photometry and the distance.  
We use the usual $\chi^2$ statistic for the likelihood, where the $\chi^2$ contributions for the light curve, the velocity curve, the spectroscopic gravity, and the computed radius are combined. In general, model parameters are drawn uniformly within specified bounds. The use of the primary radius $R_1$ as both a fitting parameter and an observed parameter effectively gives that parameter a Gaussian prior. Finally, we also include estimates of the times of the planet transits with generous uncertainties (0.0050 days for all events)
in the overall $\chi^2$. Using estimates of the transit times is very important early in the fitting process as models with transits that are very far away from the observed times would have the same $\chi^2$ contribution from the light curve as models that are much closer to the observed times since the out-of-transit parts of the light curve are flat. Including observed transit times in the overall $\chi^2$ gives a larger penalty to the models that miss the observed times by larger amounts compared to models that have smaller differences between the model transit times and the observed ones. As the models converge to the ``true'' solution, the $\chi^2$ contribution from the eclipse times becomes very small, leaving the actual fit to each individual transit profile in the light curve to determine the $\chi^2$ contribution to the overall total $\chi^2$.

The reference time for the osculating orbital parameters was chosen to be day 186 in units of BJD - 2,455,000.  The ending time of the numerical integration was day 3750 in the same units.  For these integrations we include the effects of both precession due to general relativity and precession due to tides.  As some measure of the round-off errors that afflict nearly every numerical integrator, we measure the position and velocity of the system center of mass as a function of time. These two quantities start out at {the origin [e.g.\ at the coordinates (0,0,0)]} by default, and should remain at the origin for the ideal integrator. We find that the center of mass at day 3750 is offset by 
$\approx 3.40\times 10^{-13}$ AU ($\approx 5.09$ cm), and
that the velocity at this time is offset by $\approx 2.00\times 10^{-16}$
AU per day ($\approx 3.46\times 10^{-10}$ meters per second).
Thus we conclude that numerical round-off error is not an issue for these integrations.

ELC has a number of optimizers available, and the two that proved to be the most useful for the analysis of TOI-1338 were the genetic algorithm of \citet{Charb1995} and the Differential Evolution Monte Carlo Markov Chain (DE-MCMC) algorithm of \cite{TerBraak2006}.
The spectroscopic parameters of the binary were well-known from analytic models of the radial velocities. The other parameters, especially the orbital parameters for the planet, were initially not well constrained. Thus our fitting proceeded by iteration.  The genetic code was run where an initial ``population'' of 100 models was evolved for a few thousand generations. Next, the top 80 models were evolved using the DE-MCMC code for a few thousand more generations. The top few models, randomly chosen to be between one and five, were put back into the genetic code along with random models and evolved. After a few iterations of this process, a reasonably good model was found.

After an initial good model was found, we performed several pilot runs of the DE-MCMC code. The initial population of models consisted of the best model and mutated copies of the best model where several randomly chosen parameters were offset from their optimal values by small amounts drawn from a normal distribution. The standard deviation of the normal distribution was chosen to be between 0.001 and 0.01 times the parameter value. We found that the chains quickly spread out, and achieved their final overall spreads after a burn-in period of typically 1000 generations.  During these pilot runs, we checked to confirm that the prior ranges included support 
for the entire range with non-trivial likelihood, and that
the likelihood falls to extremely small values by the 
time model parameters reach 
any of these boundaries 
(with possible exception of hard physical boundaries). We also found that the two apsidal parameters and the limb darkening coefficients for the secondary star were not constrained. Consequently, we fixed their values at representative values
($k_{2,1}=0.01$, $k_{2,2}=0.10$ for the apsidal parameters
and $q_{1,2}=0.06$, $q_{2,2}=0.41$ for the secondary's quadratic law limb darkening coefficients), reducing the number of free parameters to 21.

\begin{figure}
    \centering
    \plotone{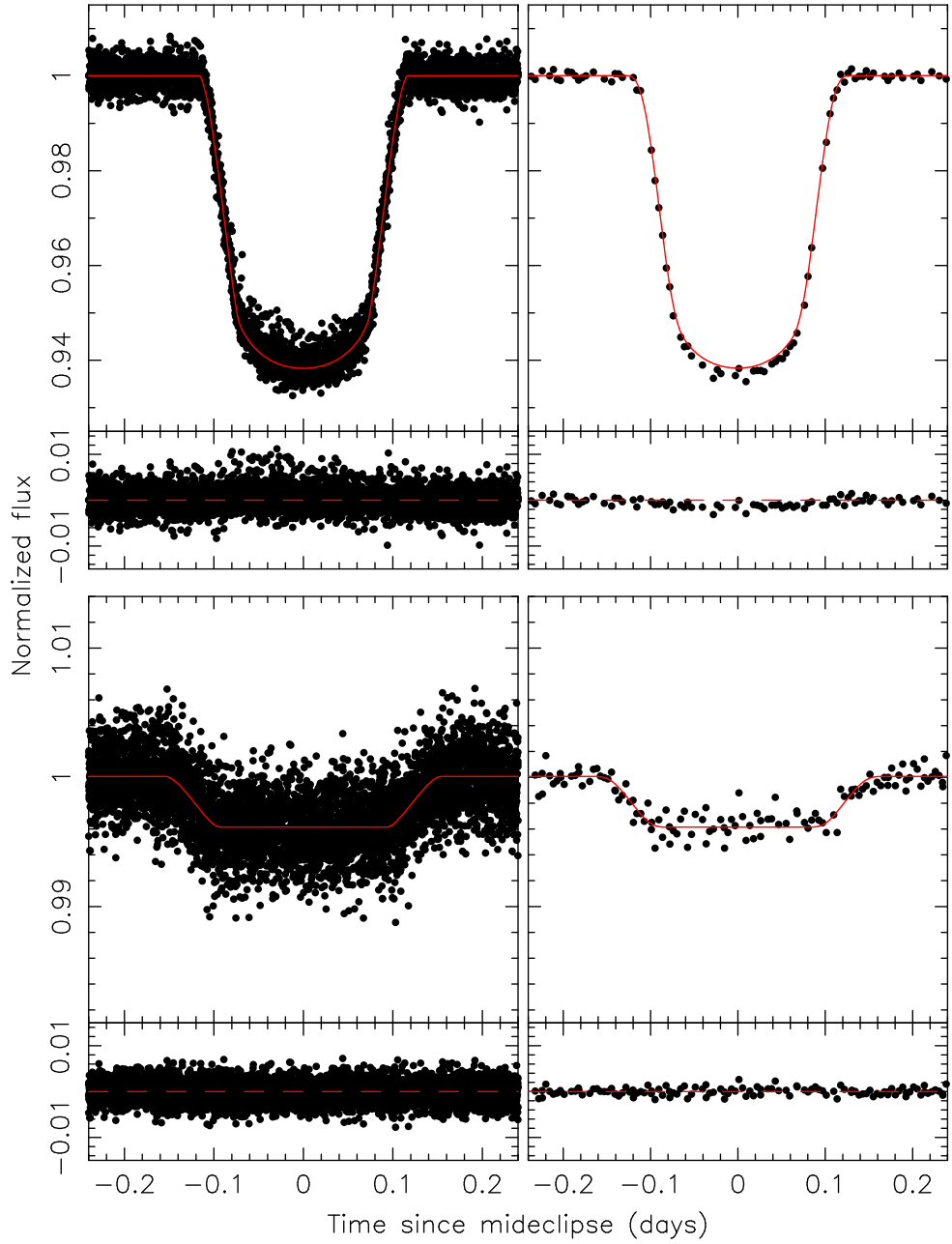}
    \caption{The folded primary and secondary eclipses are shown with the best-fitting model (solid red line). Upper left: primary eclipse in  with residuals. Upper right: Primary eclipse in and 30-min cadence with residuals. Lower left: secondary eclipse in  with residuals. Lower right: secondary eclipse in 30-min cadence with residuals. There is a slight trend of the model being shallower than the long-cadence primary eclipses.\label{fig:eclipses}}
\end{figure}

\begin{figure}
    \centering
    \plotone{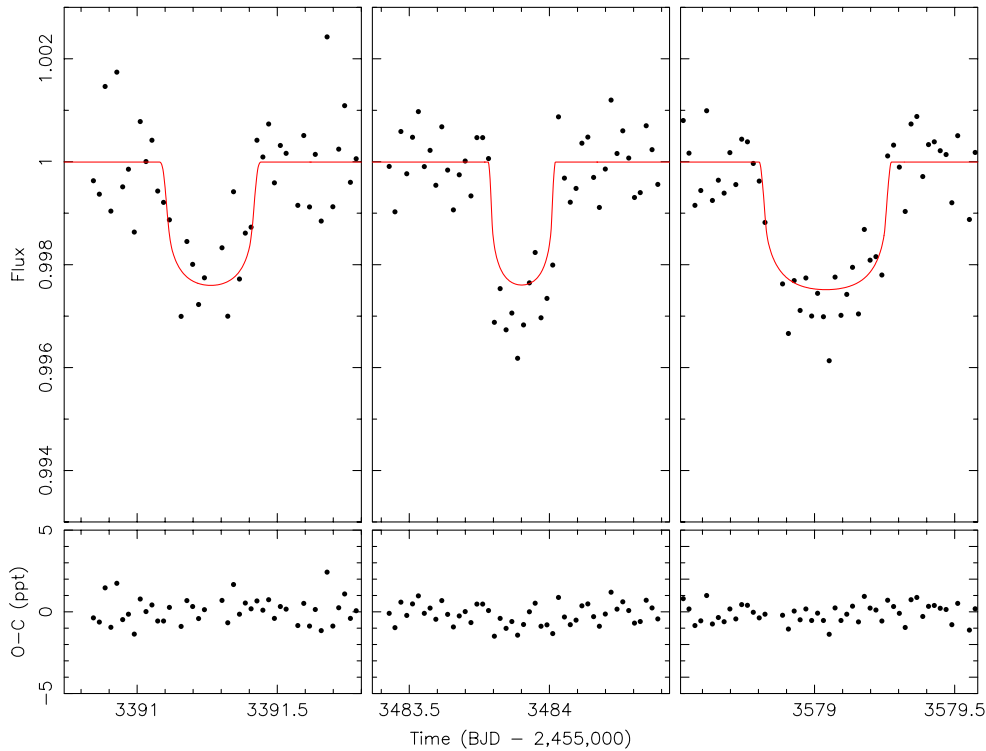}
    \caption{The transits of the CBP across the primary in Sector
    3 (left), Sector 6 (middle), and Sector 10 (right) and
    the best-fitting model.  The Sector 6 and 10 events were fit using 2-min cadence data, but for clarity we show the data binned to the 30-min cadence.\label{fig:transits}}
\end{figure}

For the final step leading to the adopted parameters and their
uncertainties, we used a brute-force ``grid search'' algorithm to find optimal models with the third body mass fixed at values from 0.3 to $117.3\,M_{\oplus}$ in steps of $1\,M_{\oplus}$ as ``seed'' models for the final runs of the DE-MCMC code.
We then ran the DE-MCMC code 8 separate times, each with the same seed models, but with a different initial random number seed. All 8 runs used 120 chains, and each was run for at least 8800 generations (the longest run had 38,300 generations). The posterior samples were drawn starting at generation 3000, with subsequent draws that skipped every 2000 generations until the chains ended.  The individual posterior samples were combined into single samples (with $N=12120$) for each fitting parameter and for several derived parameters of interest. Table \ref{tab:fitted} provides the parameters for the best-fitting model, the mode of the posterior sample (found using 50 bins), the median of the posterior sample, and the $+1\sigma$ and $-1\sigma$ uncertainties. 
Table \ref{tab:derived} provides several derived parameters of astrophysical interest using a similar format as Table \ref{tab:fitted}.  
In the discussion that follows we use the posterior medians as the adopted parameter values. The model fits to the stellar eclipses are shown in Figure \ref{fig:eclipses}, and the model fits to the planet transits are shown in Figure \ref{fig:transits}. 

There is a slight difference between the depths of the long-cadence primary eclipses and the model, such that the former are deeper. To address this issue, we reran the fit using a negative contamination factor for the long cadence data in order to make the model deeper and match the data. The best-fit dilution factor is $-0.0336\pm0.0022$, indicating that there may be a variable flux offset between the long- and short- cadence data. To first order, this suggests that either the sky background in the long cadence data was over-subtracted by 3.4\%, or the sky background in the short-cadence data was under-subtracted by 3.4\%. This dilution term seems to be somewhat on the high side of what one might expect based on the number of counts in the actual data. It is also hard to determine to which data set it has to be applied (long cadence only, short cadence only, or a combination of both). New TESS observations in Cycle 3 will help address this issue. Regardless of the specific reason and approach, the key parameters (mass, radius, etc.) do not change significantly and the results are consistent between the models with and without a dilution factor. The radial velocities are fit quite well, and Figure \ref{fig:ELCRVs} shows the residuals for the ``early'' CORALIE, the ``late'' CORALIE, and the HARPS measurements. Finally, in Table \ref{tab:Cartesian} we give the initial dynamical parameters and Cartesian coordinates for the best-fitting model to full machine precision.  

\subsection{Planet Mass}

\begin{figure}
    \hspace{-1.6cm}\includegraphics[width=0.9\textwidth,pagebox=artbox,angle=-90]{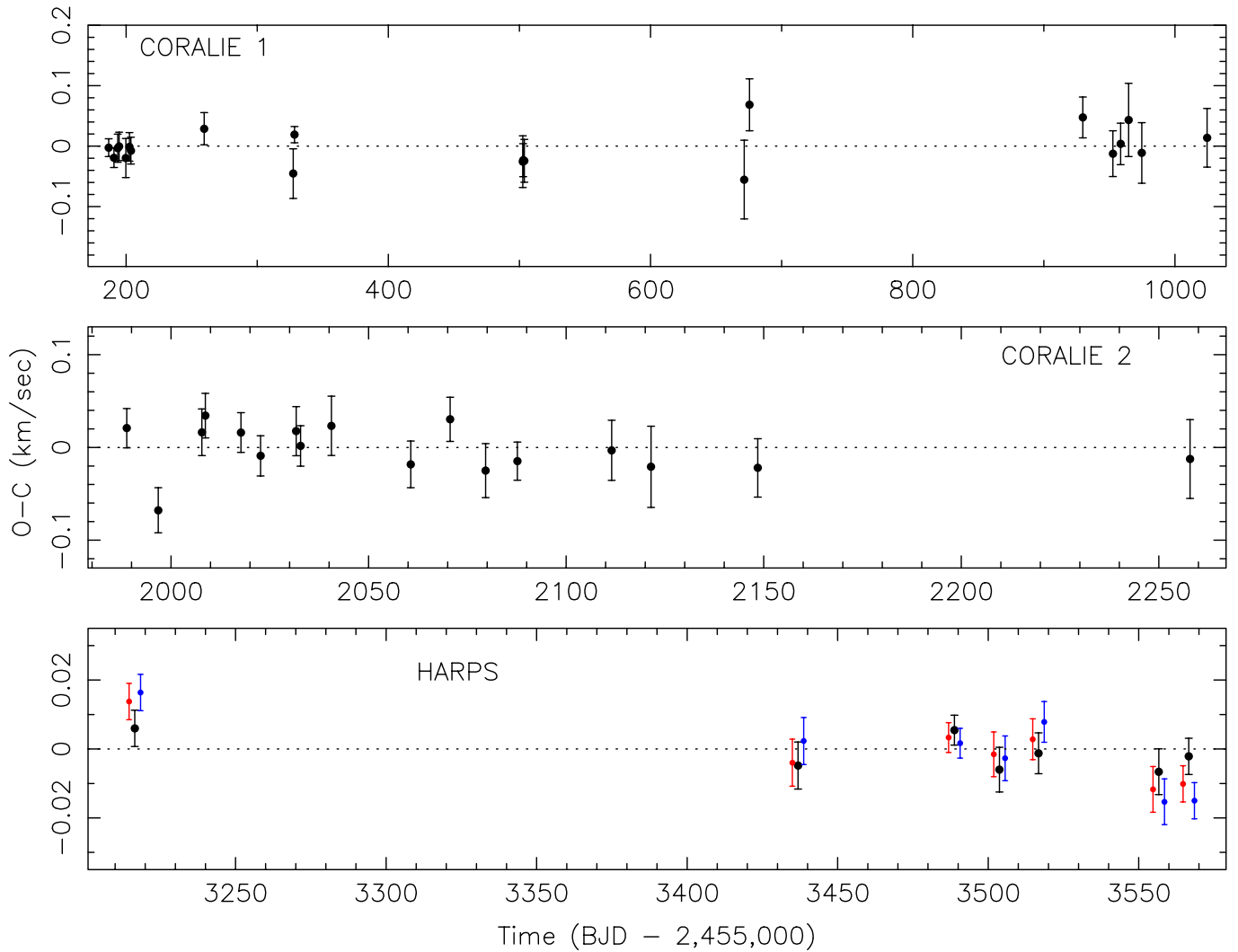}
    \caption{The residuals for the radial velocities measured by CORALIE (top and middle, black points) and HARPS (bottom, black points) for the best-fitting photodynamical model. The red and blue points show the residuals of the HARPS measurements for models with the planet mass fixed at $120\,M_{\oplus}$ and $150\,M_{\oplus}$, respectively.\label{fig:ELCRVs}}
\end{figure}

\begin{figure}
    \includegraphics[width=0.95\textwidth,pagebox=artbox,angle=0]{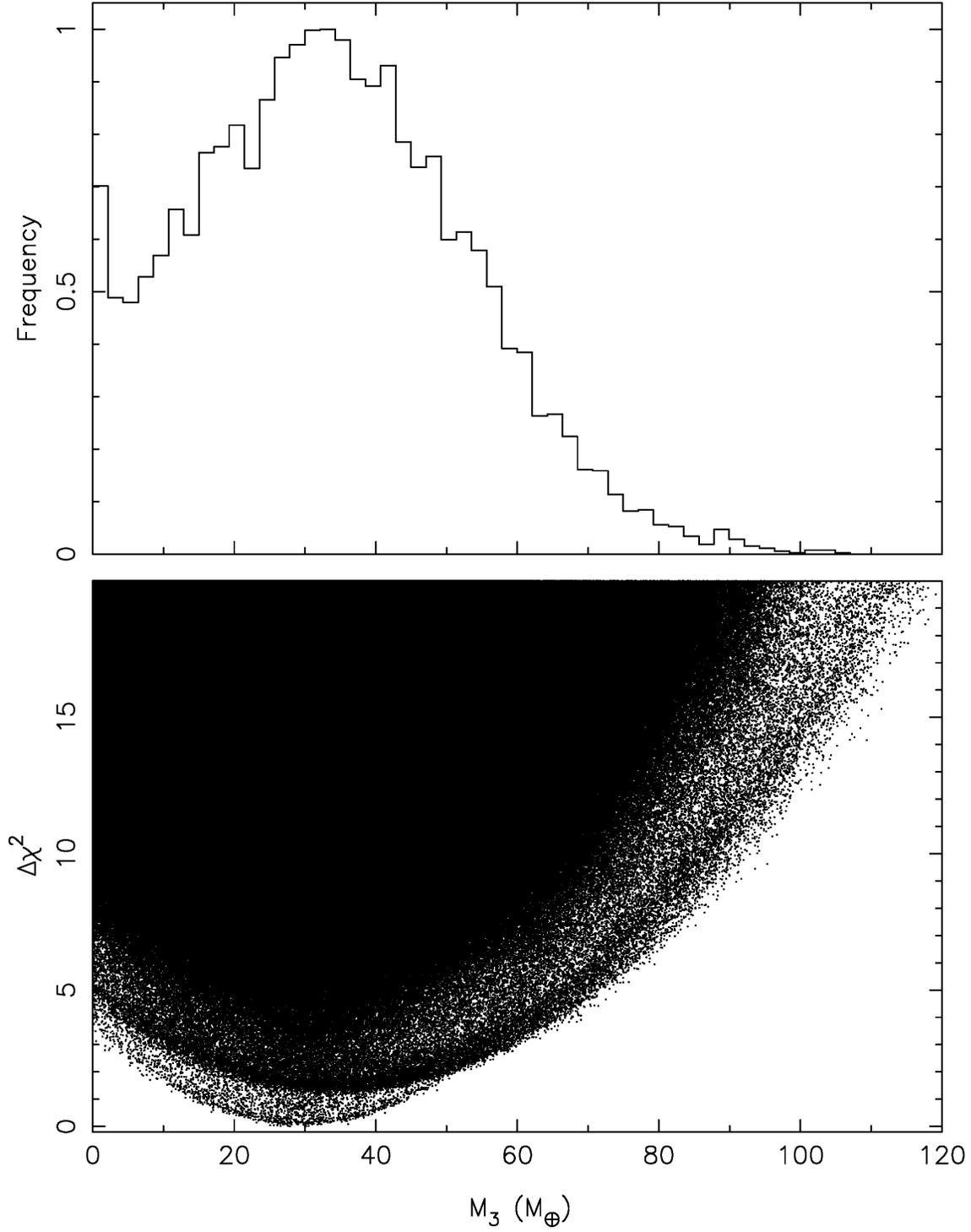}
    \caption{Top: The posterior distribution of the planet mass in units of $M_{\oplus}$.  The median value is $33.0\,M_{\oplus}$, and the largest value is $117.05\,M_{\oplus}$.  Bottom: $\chi^2-\chi^2_{\rm min}$ vs.\ the planet mass in $M_{\oplus}$ for all computed models.\label{fig:histfig01}}
\end{figure}

As noted in Section \ref{subsec:planet_discovery}, there is undoubtedly a transiting circumbinary object in the \target\ system.  Our analysis here shows that the mass of this object is well within the planetary regime, i.e.~$M_3=33\pm 20\,M_{\oplus}$. Here we give a brief discussion of what feature(s) in the data allow us to make this determination.   The top of Figure \ref{fig:histfig01} shows the posterior distribution of the planet mass $M_3$ in units of $M_{\oplus}$. The minimum and maximum values in the posterior sample are $0.01\,M_{\oplus}$ and $117.05\,M_{\oplus}$,
respectively.  The bottom of Figure \ref{fig:histfig01} shows $\chi^2-\chi^2_{\rm min}$ for all computed models, where $\chi^2_{\rm min}=14587.93$.  We see that $\Delta\chi^2$ is about 4 when $M_3=0$, and about 18 when $M_3=117\,M_{\oplus}$. As a reminder, the total $\chi^2$ has contributions from the \tess\ light curve, the three radial velocity sets, the measured gravity, and the measured radius of the primary star.  For the best-fitting model, these contributions are
14542.39 (\tess), 13.88 (early CORALIE), 17.11 (late CORALIE), 5.44 (HARPS), 8.07 ($\log g$), and 1.01 ($R_1$).  For the best model with $M_3=120\,M_{\oplus}$, those values are 14545.53 (TESS), 17.67 (early CORALIE), 20.31 (late CORALIE), 14.85 (HARPS), 9.72 ($\log g$), and 0.02 ($R_1$).  There is hardly any change in the $\chi^2$ fit to the \tess\ light curve between the two models.  The data set with the largest change in $\chi^2$ is HARPS, where the $\chi^2$ changed by about 9.4. When the planet mass is fixed at $150\,M_{\oplus}$, the total $\chi^2$ is 14626.61 (38.68 larger than the overall best model), and the individual contributions are 14553.20 (TESS), 17.96 (early CORALIE), 19.46 (late CORALIE), 25.32 (HARPS), 10.05 ($\log g$), and 0.04 ($R_1$).  Although the $\chi^2$ for the \tess\ light curve
got slightly worse, it seems that the HARPS radial velocity measurements have the most sensitivity to the planet mass.  The bottom of Figure \ref{fig:ELCRVs} shows the residuals of the HARPS measurements for the best overall model, the best model with $M_3=120\,M_{\oplus}$, and the best model with $M_3=150\,M_{\oplus}$.  The residuals of the first measurement near day 3220 and the last two measurements near days 3560 and 3570 show the most variation with the changing planet mass.  Thus in the near term the most effective way to better constrain the mass of the planet would be to obtain more radial velocity measurements with a quality similar to or better than the HARPS measurements presented here.  

\begin{figure}
    \includegraphics[width=0.95\textwidth,pagebox=artbox,angle=0]{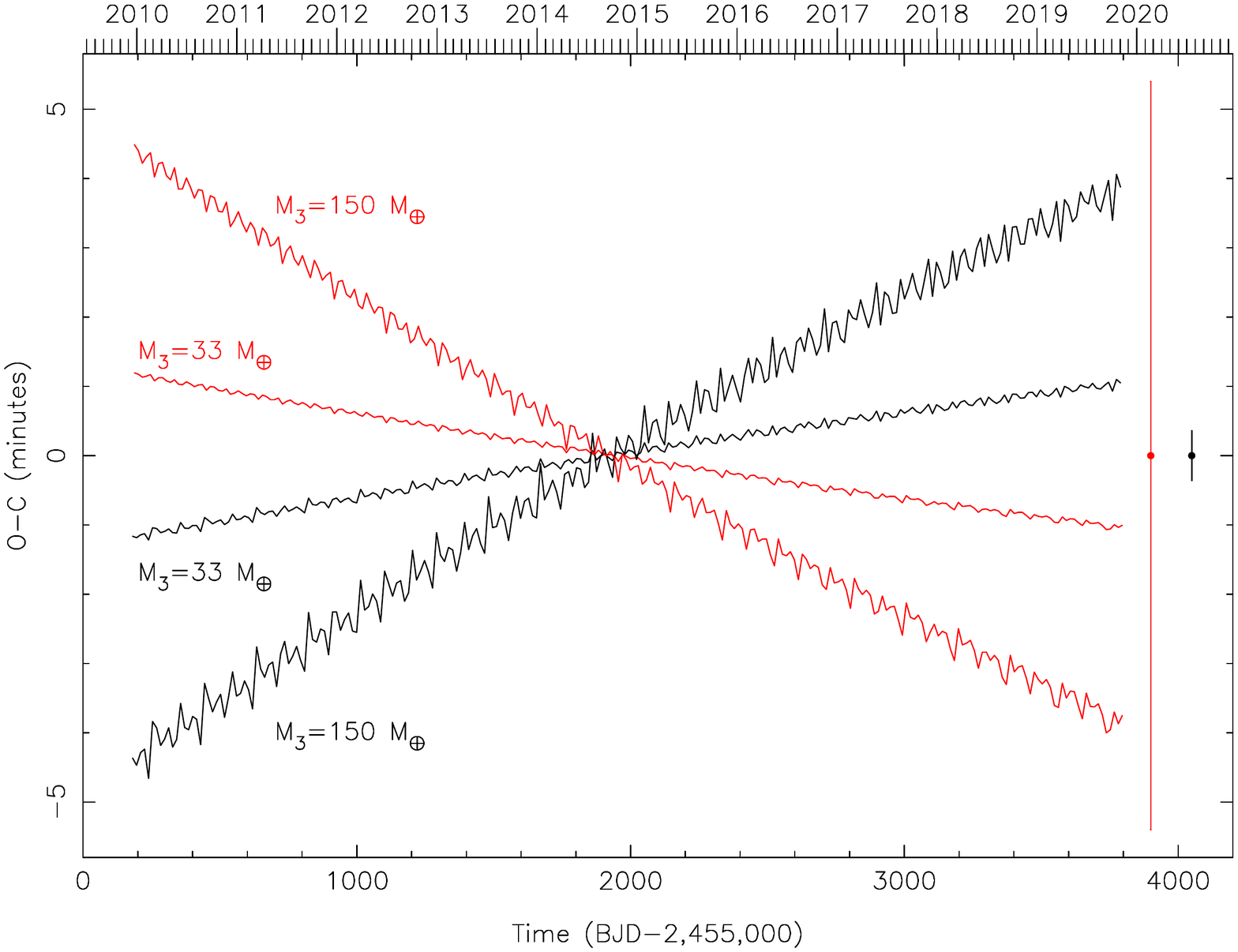}
    \caption{A Common Period O-C diagram that shows the eclipse timing variations (ETVs) in minutes for the model primary eclipses (black lines) and the model secondary eclipse (red lines). The two inner lines are for the overall best-fitting model with $M_3=33\,M_{\oplus}$, and the two outer lines
    are for the best model with the planet mass fixed at $M_3=150\,M_{\oplus}$.  The black point and error bar show the median uncertainty of 0.36 minutes for the measured times of the primary eclipses, and the red point and error bar show the median uncertainty of 5.40 minutes for the measured times of the secondary eclipses. TESS observed the system between days ${\rm \approx3336\ and\ \approx3634.}$}\label{fig:plotCPOC}
\end{figure}

A circumbinary planet can perturb the binary and give rise to eclipse time variations (ETVs).  The interaction between the planet and the binary can cause the binary orbit to precess, which leads to changes in the phase difference between the primary  and secondary eclipses.  When one attempts to fit the primary and secondary eclipse times to a common ephemeris, the O-C (Observed minus Computed) values for the primary eclipses will have the opposite slope that the O-C values for the secondary eclipses have.  Figure \ref{fig:plotCPOC} shows the Common Period O-C diagram for the model primary and secondary eclipses over the whole time span of the radial velocity and photometric observations.  For our overall best-fitting model, the argument of periastron $\omega$ changes by 0.0005715 degrees per cycle.  The contribution of this precession from General Relativity (GR) is 0.0001132 degrees per cycle, and the contribution from tides is  0.0000055 degrees per cycle.  This precession causes a divergence between the primary and secondary O-C curves of about 2 minutes over the roughly 10 year time span of the data.  When the planet mass is fixed at $150\,M_{\oplus}$, the best-fitting model for that mass has a change in $\omega$ of 0.002129 degrees per cycle.  This results in a divergence between the primary and secondary O-C curves of about 9 minutes.  As a practical matter, we only have measurements of eclipse times over the last 1.5 years or so, and the uncertainties are relatively large:  0.36 minutes for the primary eclipses and 5.40 minutes for the secondary eclipses.  Unless the measurements of the eclipse times can be vastly improved, we would need many more years of eclipse time measurements before the time baseline is long enough to accumulate a measurable divergence in the Common Period O-C diagram.

\begin{figure}
    \centering
    \plotone{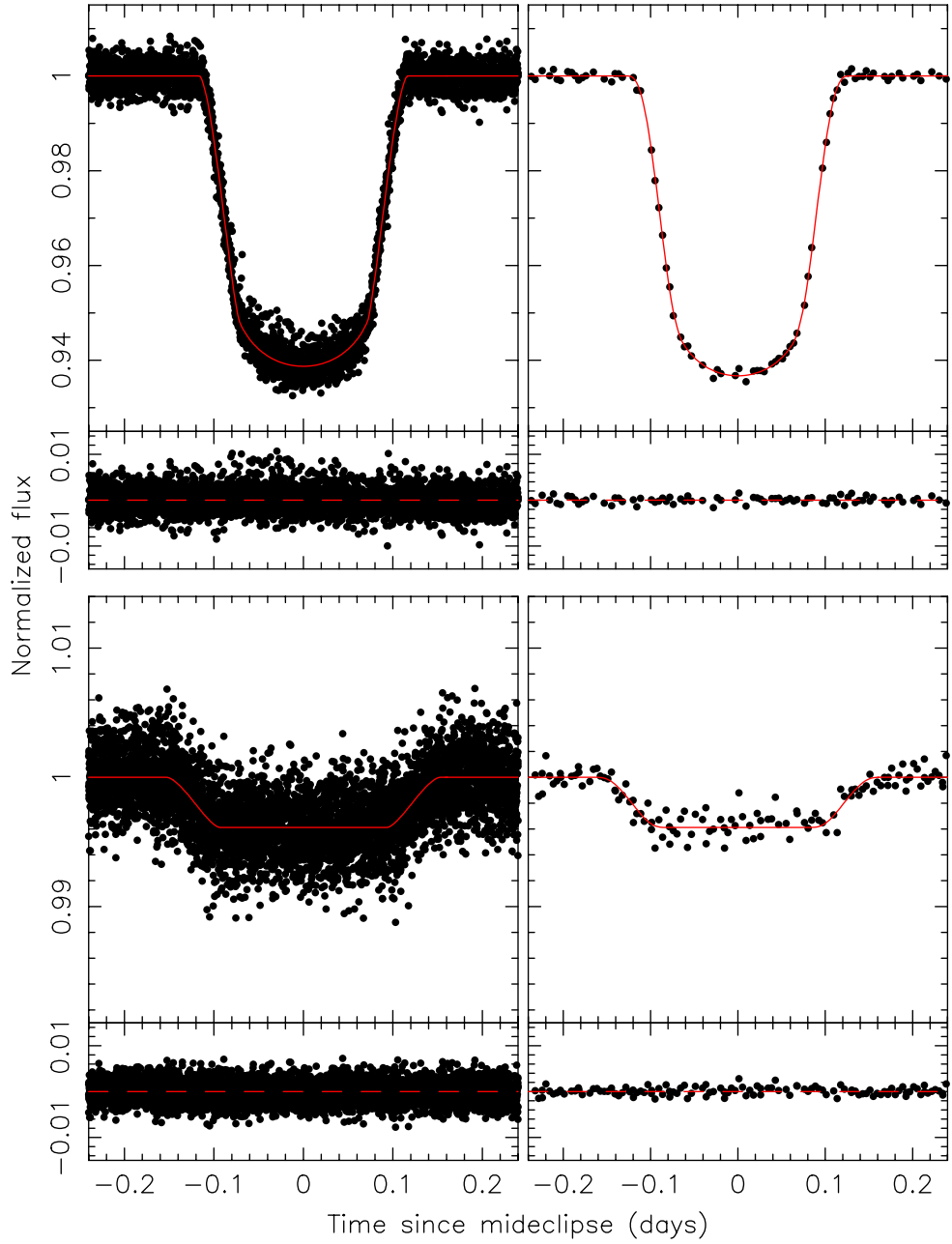}
    \caption{Same as Fig. \ref{fig:eclipses}, but for a model with a dilution factor for the long cadence data as an additional parameter to address the trend seen in that figure. While this model suggests that there is a  dilution of $\sim3.4\%$ between the long- and short-cadence data, the best-fit parameters of the system do not change significantly. \label{fig:eclipses_new}}
\end{figure}

\subsection{Dynamical Evolution}

The large tidal potential produced by the inner binary causes the orbital elements of the CBP to vary with time. Indeed, the best-fit osculating orbital elements (Table \ref{tab:derived}) represent only a snapshot at the reference epoch. These variations have consequences for both the stability and observability of CBPs (see Sect.~\ref{subsec:detectionprobabilities}). Dynamical studies of CBPs indicate that a critical regime exists, such that CBPs with periods less than ${\rm P_{crit}}$ are unstable, predominantly scattering onto an unbound orbit, or occasionally colliding with either star \citep{Dvorak1986,Holman1999,Sutherland2016,Lam2018,Quarles2018}. The process by which this instability occurs is resonant overlap \citep{Mudryk2006,Sutherland2019}. The value of $P_{\rm crit}$ is primarily a function of the dynamical mass ratio of the host stars $\mu = M_2/(M_1 + M_2)$, binary period $P_{\rm bin}$ and eccentricity $e_{\rm bin}$, but also has a dependence on the mutual inclination between the two orbits, $\Delta I$.

\begin{figure}
    \centering
    \includegraphics[width=\linewidth]{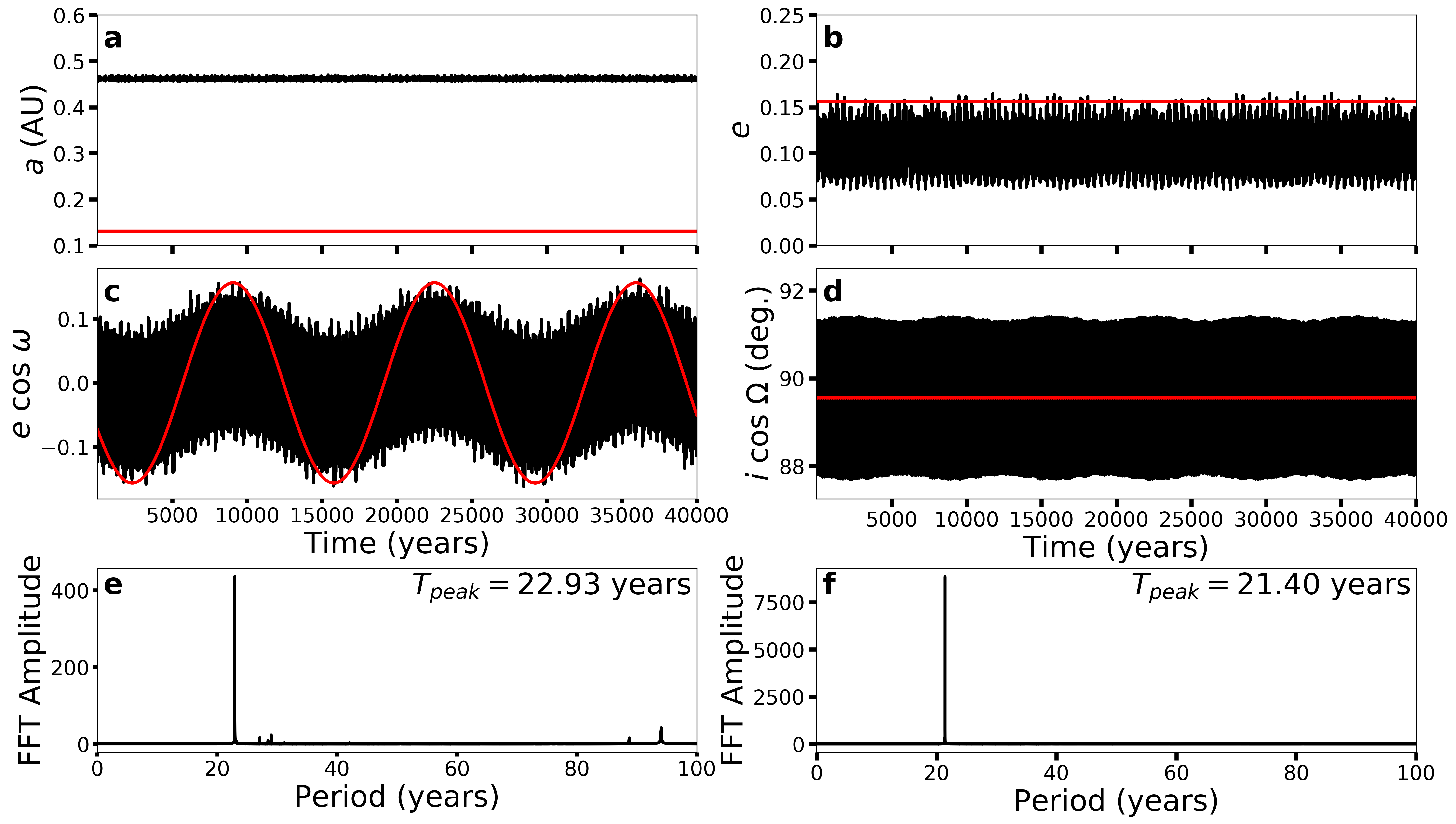}
    \caption{Evolution of the orbital elements of the CBP using the best-fit parameters from Table \ref{tab:Cartesian} for 40,000 years (BJD - 2,455,000). The top-left panel (a) shows small variations in the semimajor axis of the  planet (black) and binary (red). The top-right panel shows the variations of orbital eccentricities, indicating that the system is stable. The evolution of the $x$-components of the eccentricity (c) and inclination vectors (d) illustrate apsidal and nodal precession, respectively.  Panels (e) and (f) show Fourier periodograms using a 100,000 year simulation, where the peak values are the planetary apsidal and nodal precession periods for the respective vectors.\label{fig:TIC260128333_stab}}
\end{figure}

To investigate the dynamical evolution of the TOI-1338 system, we integrated the orbit of the CBP for $\sim$ $10^5$ yr (corresponding to $\sim{7 \times 10^3}$ orbits of
the binary) using the best-fit photometric-dynamical solution in Table \ref{tab:derived}, 
and the IAS15 integrator in the REBOUND integration package \citep{Rein2012,Rein2015}. Figure \ref{fig:TIC260128333_stab} shows the evolution of the system for 40,000 yr, using as the initial condition the best-fit photometric-dynamical solution from Table \ref{tab:Cartesian}. As seen from the figure, both the CBP (black) and binary (red) semi-major axes are practically constant with time, indicating that the system is stable. Over the course of these integrations, the eccentricity of CBP varies in a small range from 0.0695 to 0.1763 (Figure \ref{fig:TIC260128333_stab}b).

\begin{figure}
    \centering
    \includegraphics[width=\linewidth]{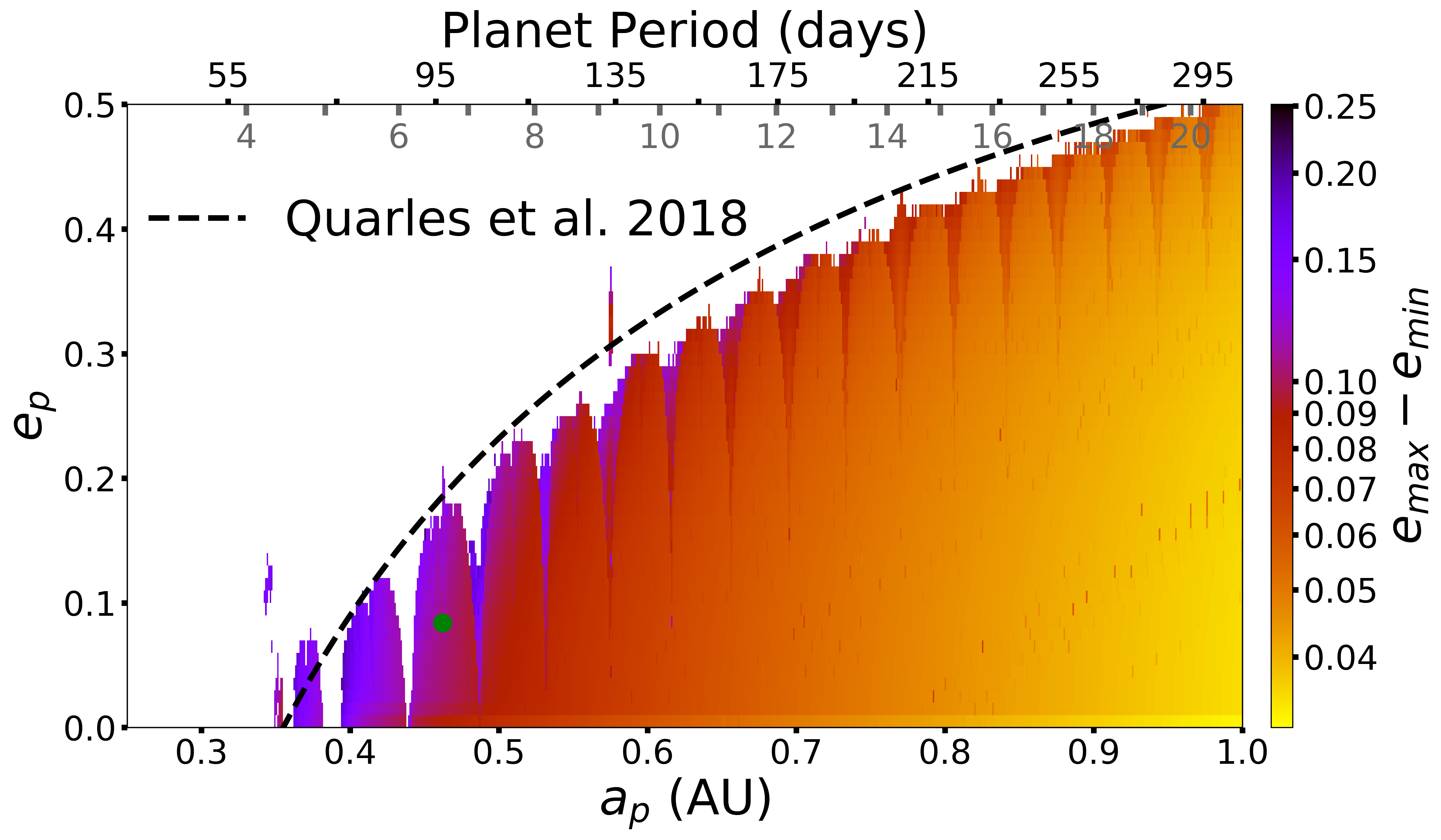}
    \caption{Stability map exploring the range of planetary eccentricity oscillation ($e_{\rm max}-e_{\rm min}$; color-coded) considering a wide range of initial values in planetary semi--major axis $a_p$ (in AU) and eccentricity $e_p$. The white cells indicate unstable initial conditions on a 100,000 yr timescale, where the vertical dips denote the locations of $N:1$ mean motion resonances (MMRs) with the inner binary. The current planetary parameters  are indicated by a green dot, where the planet lies approximately midway between the 6:1 and 7:1 MMRs. The dashed curve shows the boundary of stability from \cite{Quarles2018}, where the top axis marks the planetary period (in days) along with the location of the $N:1$ MMRs.\label{fig:TIC260128333_stab_map}}
\end{figure}

To explore whether the planet's eccentricity will continue to increase, and if that will affect the stability of its orbit, we used a modified version of the \texttt{mercury6} integration package \cite{Chambers2002} and integrated the system for $10^7$ yr. Our results showed that the extrema for the planetary eccentricity extends by only an
additional, but insignificant amount of 0.0004, confirming that the orbit of the CBP is long-term stable. Figure \ref{fig:TIC260128333_stab_map} demonstrates a more global range of stability using the \texttt{mercury6} integrator for binaries, tracking the extrema of planetary eccentricity \citep{Dvorak2004,Ramos2015}. As shown here, the orbit of the CBP (green dot) lies between 6:1 and 7:1 mean-motion resonances (MMRs)
(downward ticks, top axis) with the binary. This is important for long-term orbital stability, and is well below the eccentricity-dependent stability limit (dashed line; \cite{Quarles2018}). This is an expected result and a consequence of the fact that CBPs form at large distances away from the binary and migrate to their current orbits \citep[e.g.][]{Pierens2013,Kley2015}. Those that maintain
stable orbits are trapped between two $N:1$ MMR with the binary. This has indeed been observed in all {\it Kepler} CBPs as it is critical for long-term orbital stability. Figure \ref{fig:TIC260128333_stab_map} also shows that, although the orbit of the CBP is stable, small changes in its semimajor axis or eccentricity may 
result in a more chaotic orbit by situating the planet near a region of instability corresponding to $N:1$ MMRs with the binary. 

As an independent test to examine the stability of the planet, we used the results of our numerical integrations in the context of the scheme developed by \cite{Quarles2018}, and identified a region around the binary where the orbit of the CBP will certainly be unstable. Our analysis shows that the outer boundary of this unstable region corresponds to ${P_{crit} = 64.3}$ days (${a_{crit} = 0.36}$ AU). The observed planetary period is ${\approx50\%}$ longer than this critical value\footnote{The planetary semi-major axis is ${\approx30\%}$ larger than the critical semi-major axis.}, once again confirming that the orbit of the CBP is stable. Additionally, we also used a frequency analysis \citep{Laskar93} to obtain a quasi-periodic decomposition of the orbital perturbations of the CBP. We found these to be a combination of the five fundamental frequencies---i.e. the mean motions of the orbit of the binary, the CBP, the apsidal precession of the binary and CBP orbits, and the nodal precession---and fully consistent with the numerical simulations.

Our numerical simulations also show indications of both apsidal and nodal precessions in the orbit of the CBP. Figures \ref{fig:TIC260128333_stab}c and \ref{fig:TIC260128333_stab}d show the $x$-components of the planet's eccentricity (apsidal) and inclination (nodal) vectors. As seen here, many secular precession cycles of the planet occur within the span of 40,000 yr. The figures show a mode with a $\sim$14,286 year period, and variations that occur on a much shorter timescale of decades. We use the Fast Fourier Transform (FFT) routine within \texttt{scipy} to produce the periodograms shown in Figures \ref{fig:TIC260128333_stab}e and \ref{fig:TIC260128333_stab}f, where the system was evolved for 100,000 yr. These periodograms show strong peaks at $\sim$23 yr (8375 days) for the planetary apsidal precession period and 21.4 yr (7816 days) for the planetary nodal precession period. This nodal precession period differs slightly from the analytical result (8980 days) derived from formula given in \citet{Farago09} since the stellar binary's orbit is non-circular. 

Similar to the CBP, the orbit of the binary also experiences nodal precession. This precession is predominantly due to the perturbation of the CBP, with the tidal precession and general relativistic effect being secondary. The planetary apsidal and nodal precessions occur with similar periods but in opposite directions, as expected from \cite{Lee2006}). The longer period mode ($\sim$14,286 yr) for the planet is in phase with the binary secular precession period, where the binary causes an oscillation in the planetary argument of pericenter by $\pm$28$^\circ$ per binary precession cycle \citep{AndradeInes2018}.

\subsection{Transit and Detection Probabilities}\label{subsec:detectionprobabilities}

The nodal precession of the planet's orbit has important consequences for the long-term detectability of its transits as these can only occur when the projected path of the planet on the plane of the sky intersects with that of the stars. Nodal precession alters the planet's path, even to the extent that transits disappear for long periods of time. This was predicted by \cite{Schneider1994}, and observationally confirmed by the transits of the CBP Kepler-413b \citep{Kostov2014}. In this context, \citet{Martin2014} discussed the ``transitability'' CBPs, and found that transits of CBPs could also occur in non-eclipsing binaries.

\begin{figure}
    \centering
    \includegraphics[width=\linewidth]{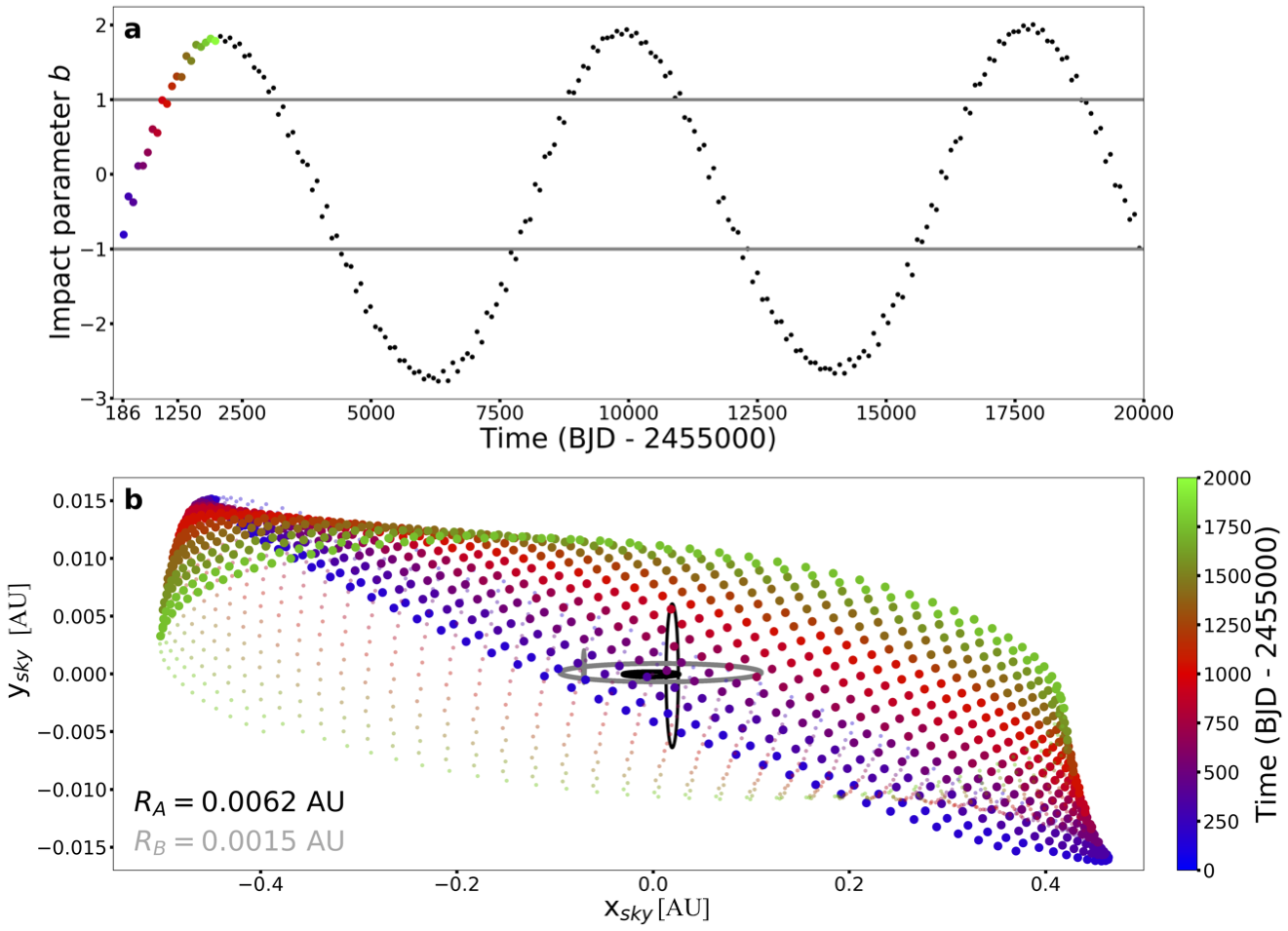}
    \caption{Evolution of the impact parameter (a) due to nodal precession of the planetary orbit for 20,000 days (BJD - 2,455,000).  Motion of the planet on the sky plane (b), where each point (color-coded) represents a small increment in the planetary orbit ($\sim$0.7 days).  The larger, opaque points indicate when the planet lies between the binary and observer, while the smaller, translucent points are when the planet lies behind the binary orbit.  The horizontal ellipses show the orbits of the primary (black) and secondary (gray) star about the center of mass, while the vertical ellipses illustrate the cross-section for transits to occur across each stellar disk.\label{fig:TIC260128333_prec}}
\end{figure}

In Figure \ref{fig:TIC260128333_prec} we show how the impact parameter of the planet, and therefore its transitability, varies over time due to the orbital evolution of the planet. As seen from the figure, there are two windows of transitability per nodal precession cycle, each roughly $\sim 1000$ days wide as indicated by the points between the horizontal gray lines in the range $-1\leq b \leq 1$. Figure \ref{fig:TIC260128333_prec}b illustrates the motion of the planet on the sky plane for the 2000 days after the starting epoch, where the points (color-coded) are spaced by $\sim 0.7$ days and vary in size (larger, opaque is towards the observer). The horizontal ellipses represent the orbits of the primary (black) and secondary (gray) stars with respect to the center of mass at the origin. The vertical ellipses indicate the cross section that each stellar disk takes up on the sky. When the larger, opaque points overlap with the vertical ellipse then transits are possible. As seen from Figure \ref{fig:TIC260128333_prec}a, the planet transits 29.7\% of the time. This is valid both for the best-fit solution from the posterior of the photometric-dynamical analysis (see Section \ref{sec:ELC}), and for the solutions taken from the overall sample from the posterior, and propagated for 155,000 days ($\sim$20 nodal precession cycles), where the planet transits $29.6\% \pm 7.6\%$ of the time. This is typical of the {\it Kepler} CBPs, for which the mean primary transitability across the first ten discovered planets was $\sim 36\%$ \citep{Martin2017}. 

\subsection{Future transits} 

As an aid to enable further observations of the TOI-1338b transits, we present in Table \ref{tab:future} the predictions of the times, impact parameters, and durations of future transit events.  These three quantities were computed using 9000 models from the posterior sample. The quoted values are the sample medians, and the quoted uncertainties are the sample r.m.s.  Transits will certainly occur on 2020 January 12, April 14, July 19, and October 19 since all 9000 models from the posterior had transits at these dates.  Starting on 2021 April 26, not all models from the posterior produce transits at that time, so the transits become less likely.  At first, the fraction of missed transits is rather small (a few percent), but then starting with the conjunction on 2023 May 20, the fraction of missed transits is large ($\sim 30\%$) and grows larger and larger thereafter.  After the 2025 September 12 conjunction, the transits have $\lesssim 30\%$ chance of occurring, and if they do occur, their impact parameters will likely be close to $-1$.

The primary \tess\ mission ends in 2020 July, but fortunately the mission has been extended through at least the year 2022\footnote{\url{https://tess.mit.edu/news/nasa-extends-the-tess-mission/}}. Depending on the exact pointing schedule, there is a good chance that \tess\ can observe transits again on 2020 October 19, 2021 January 23, 2021 April 26, and possibly July 30.

\begin{figure}
    \centering
    \includegraphics[width=0.5\textwidth]{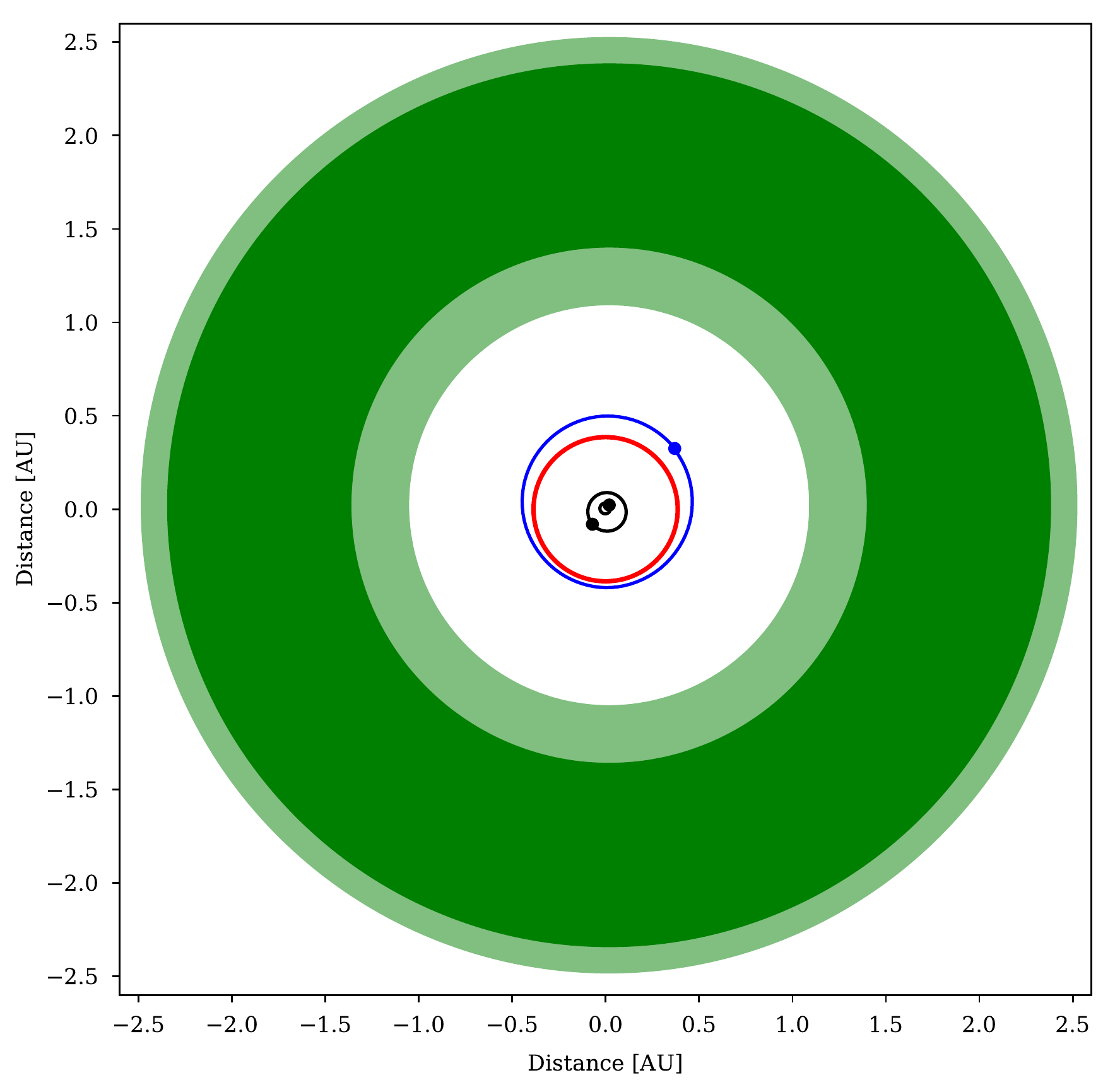}
    \caption{The extent of the system HZ, calculated based on the properties shown in Table~\ref{tab:derived}. The conservative HZ is shown in dark green and the optimistic extension to the HZ is shown in light green. The known planet is substantially interior to the boundaries of the system HZ and close to the stability boundary.\label{fig:hz}}
\end{figure}

For the sake of completeness, we also calculate the extent of the Habitable Zone \citep[HZ][]{Kasting1993} of the binary (Figure \ref{fig:hz}) using the Multiple Star Habitable Zone website developed by \citet{Muller2014}. The inner boundary of orbital stability is shown in red and the orbit of the CBP is shown in blue. TOI-1338 b is substantially interior to the HZ, receiving more than 9 times the Sun-Earth insolation.

\section{Discussion\label{sec:discussion}}

\subsection{Comparison with stellar evolution models}

\begin{figure}
    \centering
    \includegraphics[width=0.6\textwidth]{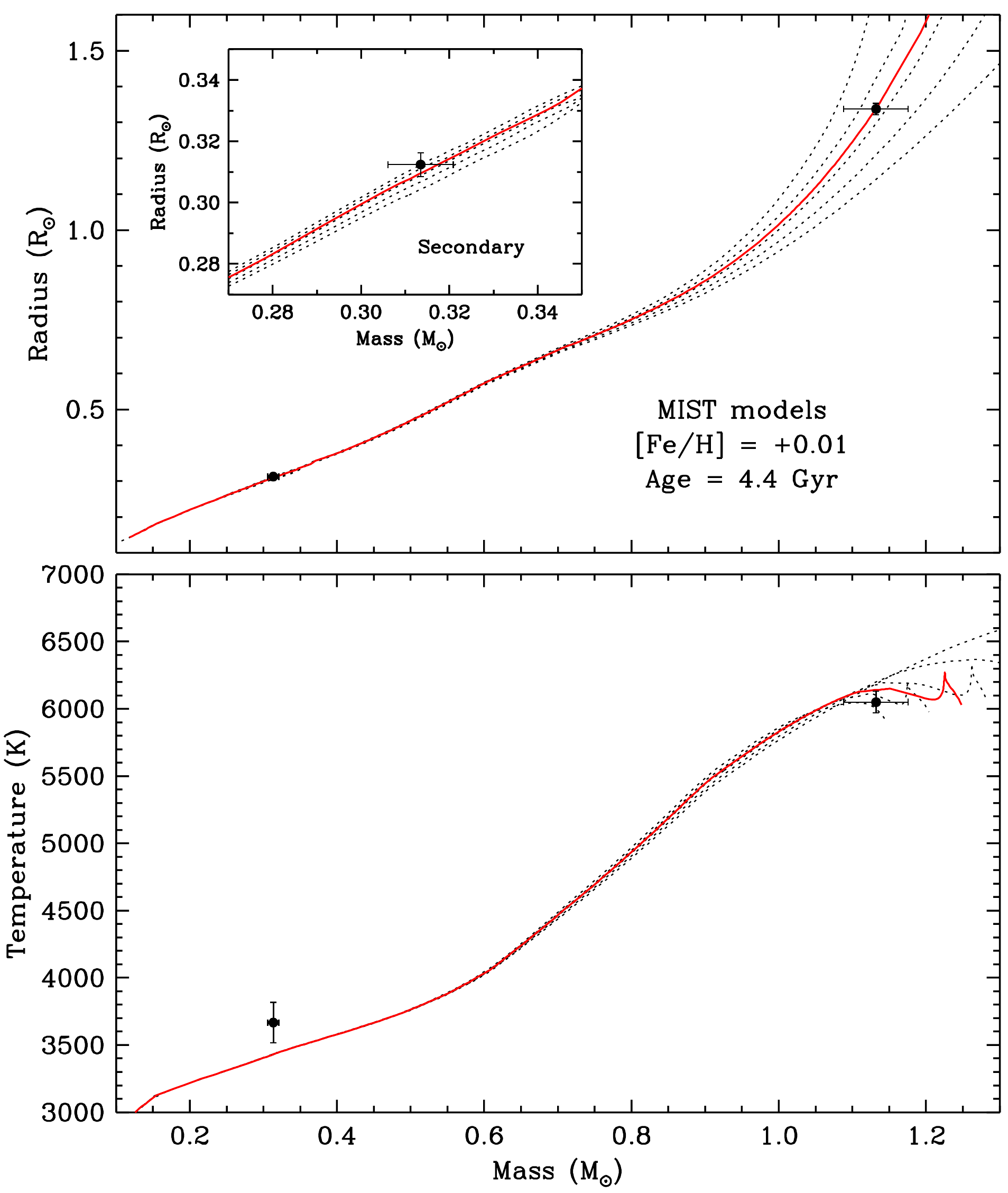}
    \caption{Upper panel: Mass and radius of TOI-1338 compared against MIST isochrones. Dotted lines represent isochrones for ages between 2 and 6 Gyr in 1 Gyr increments, and the solid line is the best-fit isochrone corresponding to 4.4 Gyr. The inset shows an enlargement around the location of the secondary star, whose properties are seen to be consistent with the best-fit model. Lower panel: Same as upper panel but for mass and temperature.\label{fig:mlr}}
\end{figure}

We compared the best-fit stellar masses, sizes and temperatures of the primary and secondary stars of TOI-1338 against model isochrones from the MIST series \citep{Choi2016} for the measured metallicity of the system. The fitted masses and sizes of both stars are in excellent agreement with a 4.4 Gyr isochrone model (see Figure \ref{fig:mlr}). It is interesting to note that the M dwarf secondary of TOI-1338 does not seem to be significantly inflated compared to standard models, as may be the case for some CBP systems with similar secondary stars 
\citep[e.g.\ Kepler-38 and Kepler-47,][]{Orosz2012,Orosz2019}. However, this is in line with the rest of objects detected during the EBLM Project, which show no systematic radius inflation for fully-convective low mass stars \citep{Boetticher2019}. The effective temperature of the secondary star is also consistent with model predictions, within the errors.

%
%

\subsection{Stellar rotation}

Starspots create modulations in the light curves of eclipsing binaries (typically seen in the out-of-eclipse regions), which affect measurements of eclipse times and thus photodynamical models (see e.g. Kepler-453, Welsh et al. 2018). The systematic errors present in the \tess\ light curve of TOI-1338 (see Fig. \ref{fig:LC_fig_full}) preclude measurement of the intrinsic stellar variability from the \tess\ data alone which, in turn, precludes determination of the stellar spin period based on starspot modulations. 

\begin{figure}
    \includegraphics[width=0.9\textwidth]{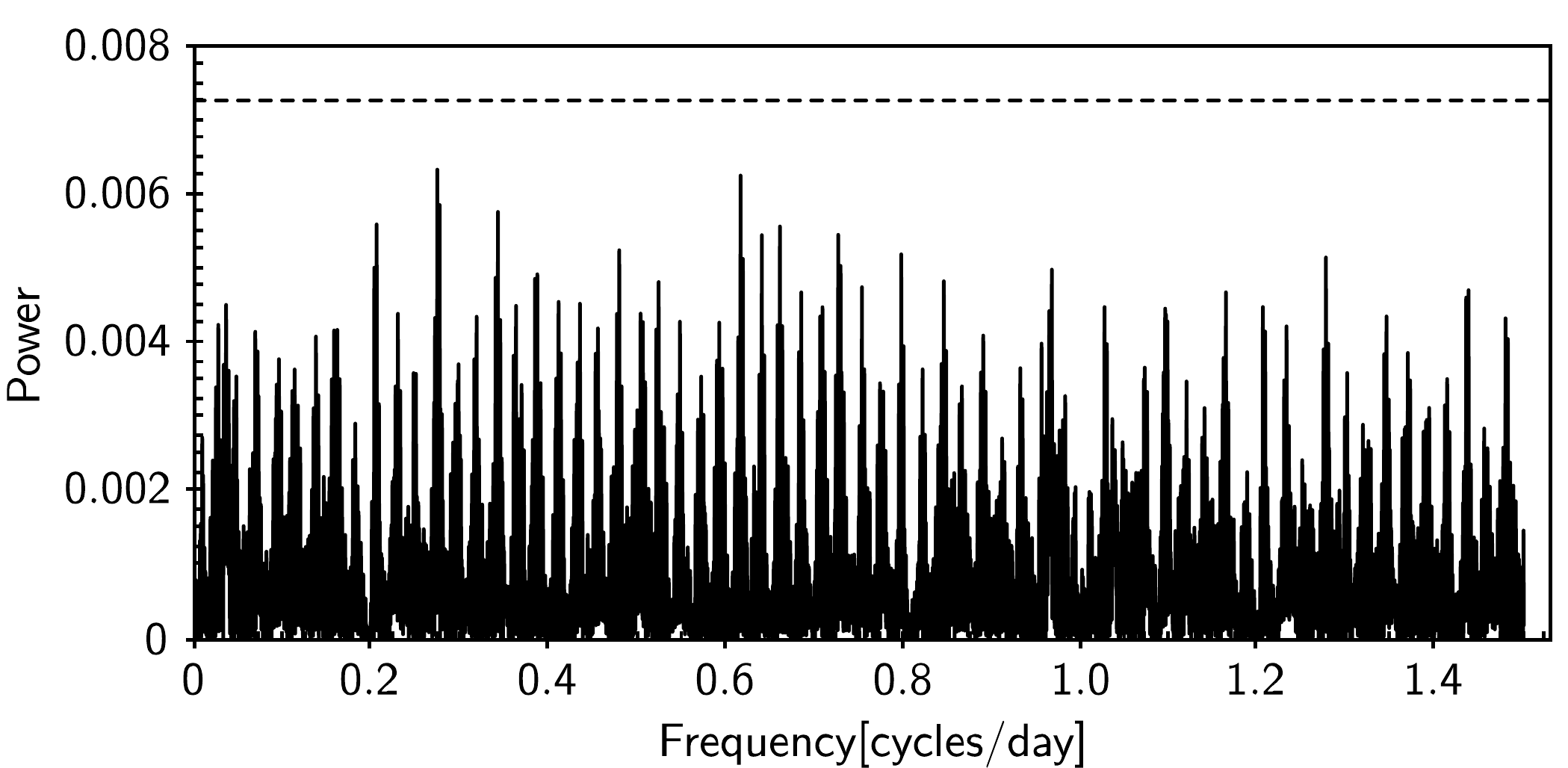}
    \caption{Periodogram of the WASP lightcurve for TIC~260128333. The dashed line indicates a false-alarm probability level of 0.01.\label{fig:EBLM0608-59_222_4738-5646_pgram}}
\end{figure}

However, if we assume that the stellar rotation axis is approximately perpendicular to the line of sight, then our measurement of $v\,\sin i_{\star}$ combined with the stellar radius estimated above imply a rotation period for the primary star of $P_{\rm rot} = 19\pm3$\,days. Given that the binary orbit is eccentric, and that the timescale for synchronization (${\rm \sim 2.5}$ Gyrs) is comparable to the estimated age (${\rm \sim 4.4}$ Gyrs), we can expect the rotation period to be closer to the pseudosynchronous period (${\sim12.7}$ days). If this star is magnetically active then we expect to see modulation of the light curve at frequencies $1/P_{\rm rot}$ and/or $2/P_{\rm rot}$, depending on the distribution of active regions on the stellar surface at the time of observation. To search for such modulations and periodic signals in the WASP light curve of TOI-1338, we used the sine-wave fitting method described in \citet{Maxted2011}. The WASP light curve contains 26,492 observations obtained with the same CCD camera and 200-mm lens over 3 observing seasons. We calculated the periodogram over 32,768 uniformly spaced frequencies from 0 to 1.5 cycles/day. The false alarm probability (FAP) is calculated using a boot-strap Monte Carlo method also described in \citet{Maxted2011}. The periodogram is shown in Fig.~\ref{fig:EBLM0608-59_222_4738-5646_pgram}. From the boot-strap Monte Carlo simulations and the lack of any significant signal in this periodogram we can put an upper limit of approximately 1\,mmag to the semi-amplitude of any signal due to rotational modulation. A similar analysis for the 3 seasons of WASP data separately is consistent with this conclusion. 

\begin{figure}
    \centering
    \includegraphics[width=0.9\textwidth]{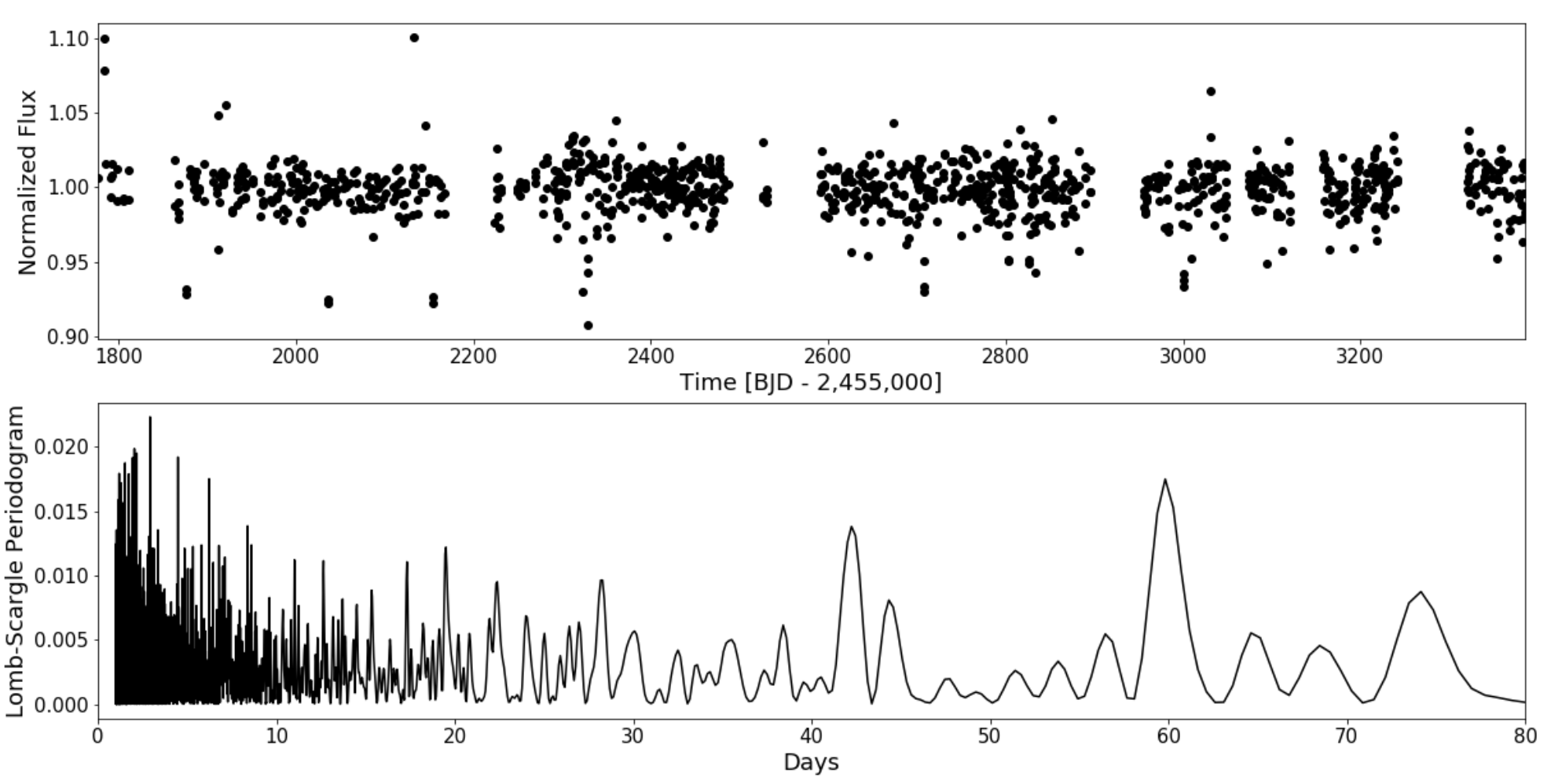}
    \caption{Approximately 1600 days of ASAS-SN V-band photometry showing no significant rotationally-induced photometric modulation.\label{fig:ASAS_SN}}
\end{figure}

To further this point, we also obtained historical data from the All-Sky Automated Survey for Supernovae (ASAS-SN) project \citep{Shappee14, Kochanek17} in order to search for evidence of star spot modulation. Our analysis indicates that ${\sim 1600}$ days of ASAS-SN V-band data shows no significant photometric modulation either (Fig. \ref{fig:ASAS_SN}). Additionally, the HARPS spectra have a dispersion of only a few m~s$^{-1}$, compatible with a chromospherically quiet star. The bisectors of the cross-correlation function likewise show no variability \citep{2001A&A...379..279Q}. These considerations further strengthen our assumption that the stellar activity and starspot-induced bias in the measured eclipse times is negligible, and that their effect on the photodynamical solution is minimal. If additional photometric observations should reveal the primary star to be heavily spotted, then the planet mass determination may need to be revised.

\subsection{The Planet}

With a mass of $33.0\pm 20.0\,M_{\oplus}$, a radius of
$6.85\pm 0.19\,R_{\oplus}$, and a bulk density of
$0.56\pm 0.34$ g cm$^{-3}$ (Table \ref{tab:derived}, where
the quoted values are from the sample medians), the closest
Solar System analogue of \target b is perhaps Saturn, where $M=95.16\,M_{\oplus}$,
$R=9.14\,R_{\oplus}$, and $\rho=0.69$ g cm$^{-3}$. Among the known
CBPs, \target b has bulk properties similar to those of Kepler-16b \citep{Doyle2011} and Kepler-34b \citep{Welsh2012}.

The large radius of \target b is consistent with the predictions of planet-formation models in circumbinary disks, a trend that has also been observed among CBPs detected by the {\it Kepler} telescope. Combined with the small orbital inclination, this indicates that \target b formed at larger distances from its host binary and migrated to its  current orbit through planet-disk interaction  \citep[e.g.\ ][]{Pierens2013,Kley2015}. 

\subsection{TOI-1338 within the context of the 
{\em Kepler} CBP systems}

\begin{figure}
    \centering
    \includegraphics[width=0.9\textwidth]{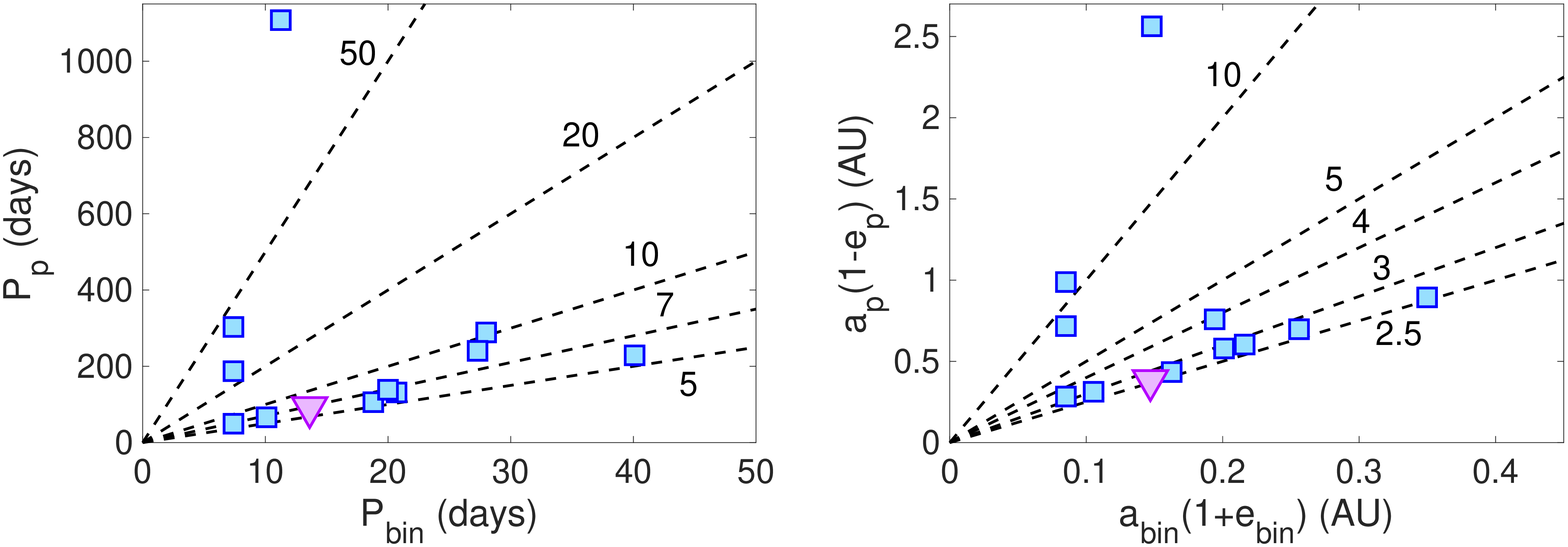}
    \caption{(a) Ratio of the planet and binary orbital periods for Kepler CBPs (blue squares) and \tess\ (purple downwards triangle). The diagonal black dashed lines indicate constant period ratios. (b) Corresponding ratio of the planet periapse and the binary apoapse, which together represent the shortest separation between the planet and binary.\label{fig:period_semi_ratios}}
\end{figure}

The TOI-1338 system follows the trends established by {\em Kepler} CBPs. Namely, these are gas giant planets (radius larger than $3R_{\oplus}$) with low-eccentricity, nearly co-planar orbits\footnote{${\rm e_{CBP}~<~0.15}$, mutual orbital inclination ${\rm \Delta i < 5^{\circ}}$.} with periods longer than ${\sim 50}$ days, and orbit around binary stars with $P_{\rm bin}\sim 7.5-40$ days \citep[][and references therein]{Welsh2018,Martin2019}\footnote{Whereas most of {\it Kepler's} EBs have orbital periods shorter than 
${\sim 3}$ days \citep{Munoz2015,Martin2015,Hamers2016,Fleming2018}.}. 
The near co-planarity of the planetary orbits is also consistent with the observational results of circumbinary disks around short period stellar biaries \citep{Czekala19}. \target\ is similar to the Kepler-38 CBP system \citep{Orosz2012a} in terms of both the orbital periods and orbital period ratio (see Figure \ref{fig:period_semi_ratios}). The orbital precession timescale of \target b is comparable to that of the CBP Kepler-413b, where the time scale is $\sim 22$ years for the former compared to $\sim 11$ years for the latter \citet{Kostov2014}.

\begin{figure}
    \centering
    \includegraphics[width=0.9\textwidth]{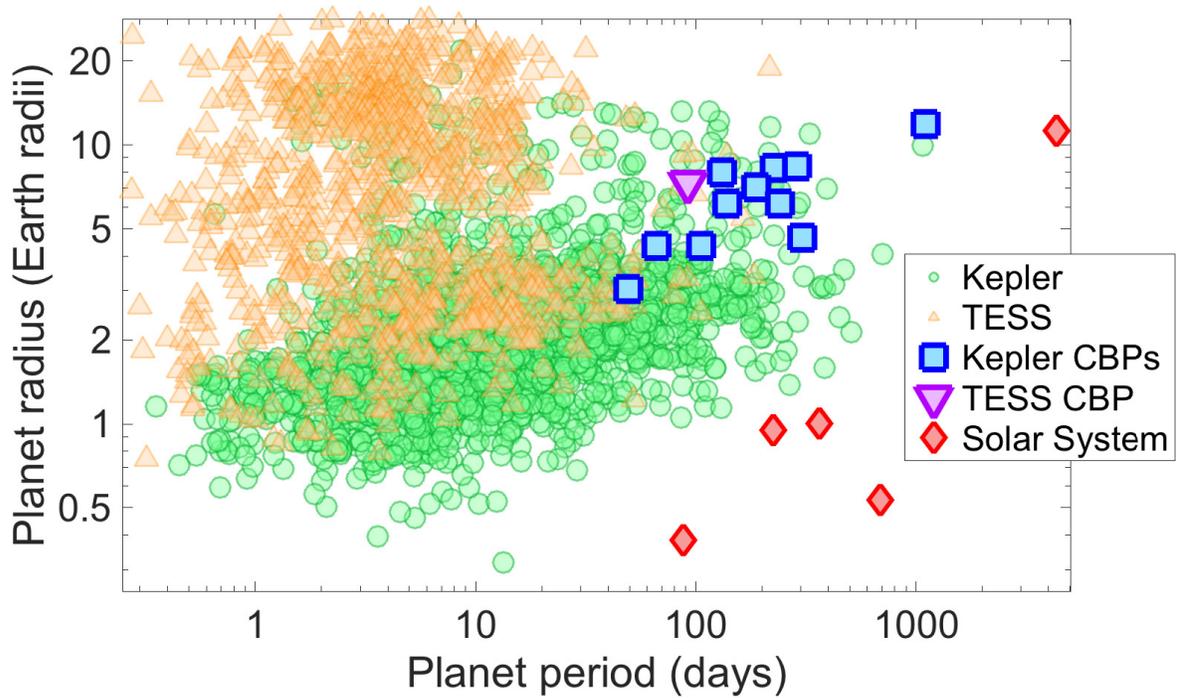}
    \caption{Radius and period of the \tess\ circumbinary planet (purple downwards triangle) and Kepler circumbinary planets (blue squares) compared with the \tess\ planet candidates (yellow upwards triangles), Kepler planet candidates (green circles) and the 5 innermost Solar System planets (red diamonds). This circumbinary discovery is the longest-period confirmed \tess\ planet to date, and exists in a parameter space similar to the Kepler circumbinary planets.\label{fig:period_radius}}
\end{figure}

Overall, TOI-1338b has a relatively long period for a transiting planet, particularly when compared with other \tess\
candidates. This is demonstrated in Figure~\ref{fig:period_radius}, where it resides on the very tail of the \tess\ 
planet candidate period distribution. We note that the current lack of small CBPs is likely an observational bias, since unique challenges have inhibited their detection to date, largely as a result of the transit timing variations induced by the barycentric binary motion and the orbital dynamics (Armstrong et al. 2013, Armstrong et al. 2014, Martin 2019a, Windemuth et al. 2019). The circumbinary planet population is yet to be constrained below $4R_{\oplus}$ (Armstrong et al. 2014), and we expect that the large quantity and brightness of the \tess\ stars will enable the expansion of this parameter space.

\section{Conclusions\label{sec:end}}

We presented the discovery of the first transiting circumbinary planet from \tess, 
\target. The target was observed by \tess\ in 12 sectors of 30-min cadence data (Sectors 1 through 12), and 9 sectors of 
2-min cadence data (Sectors 4 through 12).  In addition to stellar eclipses,
three transits events were observed in Sectors 3, 6, and 10. These extra transit events show the hallmark characteristics of a circumbinary object where their duration depends on the binary phase and their times have significant deviations from a simple linear ephemeris.  Blending is not an issue with \target\ as the nearest source is $53\arcsec$ away, and speckle imaging observations from SOAR rule out nearby sources with a magnitude difference of $\Delta I\sim 4$ down to $0.5\arcsec$ from the target.
Radial velocity measurements are available as the host eclipsing binary has been monitored for more than three years by CORALIE and HARPS as part of the EBLM project. 
To solve for the parameters of the system, we combined the \tess\ data with the radial velocities into the photometric-dynamical model ELC.
Our analysis confirms that the circumbinary object is indeed a planet,
with a mass of $33.0\pm 20.0\,M_{\oplus}$, a radius of
$6.85\pm 0.19\,R_{\oplus}$, and a bulk density of
$0.56\pm 0.34$ g cm$^{-3}$.  The planet's orbit is within 
${\sim 1^{\circ}}$ of being coplanar with the binary,
has a period of $95.2$ days and small eccentricity, and is safely beyond the boundary for stability. The host eclipsing binary (with $P=14.6$ days and $e\approx 0.16$) 
consists of G+M stars with masses $1.1\,M_{\odot}$ and $0.3\,M_{\odot}$, and radii of
$1.3\,R_{\odot}$ and $0.3\,R_{\odot}$ respectively. Based on the stellar parameters, we estimate an age of 4.4 Gyr for the system.

\acknowledgments

This manuscript includes data collected by the \tess\ mission, which are publicly available from the Mikulski Archive for Space Telescopes (MAST). Funding for the TESS mission is provided by NASA Science Mission directorate. We acknowledge the use of public \tess\ data from pipeline at the \tess\ Science Processing Operations Center. The manuscript includes data from CORALIE, an instrument mounted on the Euler 1.2m telescope, a project of the University of Geneva, funded by the Swiss National Science Foundation. Furthermore, our analysis includes spectra obtained with HARPS, an instrument mounted on the ESO-3.6m telescope at La Silla. Those data were obtained under Prog.ID 1101.C-0721 (PI Triaud). They are or will become available through the ESO public archive.
Resources supporting this work were provided by the NASA High-End Computing (HEC) Program through the NASA Advanced Supercomputing (NAS) Division at Ames Research Center for the production of the SPOC data products.  This research was supported in part through research cyberinfrastructure resources and services provided by the Partnership for an Advanced Computing Environment (PACE) at the Georgia Institute of Technology.  This research made use of Lightkurve, a Python package for Kepler and \tess\ data analysis (Lightkurve Collaboration, 2018). This research has made use of the Exoplanet Follow-up Observation Program website, which is operated by the California Institute of Technology, under contract with the National Aeronautics and Space Administration under the Exoplanet Exploration Program. We are also grateful to the observer support staff at CTIO, at ESO/HARPS and Swiss Euler Telescope/CORALIE. This work has made use of data from the European Space Agency (ESA) mission Gaia (https://www.cosmos.esa.int/gaia), processed by the Gaia Data Processing and Analysis Consortium (DPAC, https://www.cosmos.esa.int/web/gaia/dpac/consortium). Funding for the DPAC has been provided by national institutions, in particular the institutions participating in the Gaia Multilateral Agreement. WFW and JAO thank John Hood, Jr. for his generous support of exoplanet research at SDSU.  Support was also provided and acknowledged through NASA Habitable Worlds grant 
80NSSC17K0741 and NASA XRP grant 80NSSC18K0519. This work is partly supported by NASA Habitable Worlds grant 80NSSC17K0741. This material is based upon work supported by the National Science Foundation Graduate Research Fellowship Program under Grant No. (DGE-1746045). AHMJT has received funding from the European Research Council (ERC) under the European Union's Horizon 2020 research and innovation programme (grant agreement n${^\circ}$ 803193/BEBOP), and from a Leverhulme Trust Research Project Grant n${^\circ}$ RPG-2018-418. AC acknowledges support by CFisUC strategic project (UID/FIS/04564/2019). Any opinions, findings, and conclusions or recommendations expressed in this material are those of the authors and do not necessarily reflect the views of the National Science Foundation. E.A.G. thanks the LSSTC Data Science Fellowship Program, which is funded by LSSTC, NSF Cybertraining Grant \#1829740, the Brinson Foundation, and the Moore Foundation; her participation in the program has benefited this work. E.A.G. is thankful for support from GSFC Sellers Exoplanet Environments Collaboration (SEEC), which is funded by the NASA Planetary Science Division’s Internal Scientist Funding Model.

\vspace{5mm}
\facilities{\tess; SOAR~4.0-m; WASP; Swiss Euler Telescope (CORALIE); ESO~3.6-m (HARPS)}

\software{astropy \citep{2013A&A...558A..33A}; scipy \citep{Jones2019}; eleanor \citep{feinstein2019}; Lightkurve (Lightkurve Collaboration et al. 2018); ispec (Blanco-Cuaresma et al. 2014); REBOUND (Rein \& Liu 2012; Rein \& Spiegel 2015); Mercury6 (Chambers et al. 2002)
          }

\clearpage

%
%
\begin{deluxetable}{lrrr|lrrr}
\tablecaption{Observed eclipse times for \target\label{eclipsetimes1}}
\tablewidth{0pt}
\tablehead{
\colhead{Cycle}   & 
\colhead{Observed Time\tablenotemark{a}}  &
\colhead{Model Time\tablenotemark{a}}    &
\colhead{O-C\tablenotemark{b}}  &
\colhead{Cycle} &
\colhead{Observed Time\tablenotemark{a}} &
\colhead{Model Time\tablenotemark{a}}  &
\colhead{O-C\tablenotemark{b}}       }
\startdata
\multicolumn{4}{c|}{{Primary}} & \multicolumn{4}{c}{{Secondary}} \\
\hline
  0  &    \nodata                   & 3322.21319 & \nodata & 0.45345 & $3328.83160 \pm0.00557$ & 3328.83833 & $-9.71$ \\
  1  &    $3336.82137\pm   0.00027$ & 3336.82171 & $-0.48$ & 1.45345 & $3343.44848 \pm0.00970$ & 3343.44687 & $2.32$  \\
  2  &    $3351.43036\pm   0.00026$ & 3351.43028 & $0.12$  & 2.45345 & $3358.05700 \pm0.00285$ & 3358.05537 & $2.34$  \\
  3  &    $3366.03902\pm   0.00023$ & 3366.03888 & $0.21$  & 3.45345 & $3372.65843 \pm0.00410$ & 3372.66390 & $-7.87$ \\ 
  4  &    $3380.64698\pm   0.00059$ & 3380.64735 & $-0.52$ & 4.45456 & $3387.27302 \pm0.00379$ & 3387.27251 & $0.73$ \\
  5  &    $3395.25589\pm   0.00044$ & 3395.25598 & $-0.12$ & 5.45345 & $3401.86970 \pm0.01653$ & 3401.88102 & $-16.29$ \\
  6  &    \nodata                   & 3409.86452 & \nodata & 6.45345 & $3416.49508 \pm0.00375$ & 3416.48957 & $7.93$  \\
  7  &    \nodata                   & 3424.47302 & \nodata & 7.45345 & $3431.09816 \pm0.00340$ & 3431.09811 & $0.06$  \\
  8  &    $3439.08165\pm   0.00020$ & 3439.08157 & $0.13$  & 8.45345 & $3445.70014 \pm0.00459$ & 3445.70662 & $-9.34$ \\
  9  &    $3453.69010\pm   0.00019$ & 3453.69017 & $-0.09$ & 9.45345 & $3460.32094 \pm0.00364$ & 3460.31513 & $8.36$ \\
 10  &    \nodata                   & 3468.29871 & \nodata & 10.45345& $3474.92301 \pm 0.00324$ & 3474.92372 & $-1.02$ \\
 11  &    $3482.90704\pm   0.00022$ & 3482.90721 & $-0.24$ & 11.45345& $3489.53315 \pm 0.00300$ & 3489.53225 & $1.30$ \\
 12  &    $3497.51571\pm   0.00021$ & 3497.51582 & $-0.15$ & 12.45345& \nodata                 & 3504.14078 & \nodata \\
 13  &    $3512.12410\pm   0.00022$ & 3512.12432 & $-0.31$ & 13.45345& $3518.73613 \pm0.00491$ & 3518.74934 & $-19.02$\\
 14  &    $3526.73304\pm   0.00019$ & 3526.73285 & $0.29$  & 14.45345& \nodata                 & 3533.35786 & \nodata \\
 15  &    $3541.34140\pm   0.00022$ & 3541.34142 & $-0.02$ & 15.45345& $3547.97016 \pm0.00444$ & 3547.96636 & $5.48$ \\
 16  &    \nodata                   & 3555.95001 & \nodata & 16.45345& $3562.57901 \pm0.00349$ & 3562.57490 & $5.91$ \\
 17  &    \nodata                   & 3570.55847 & \nodata & 17.45345& $3577.18768 \pm0.00349$ & 3577.18350 & $6.02$ \\
 18  &    $3585.16712\pm   0.00032$ & 3585.16712 & $0.02$  & 18.45345& $3591.78979 \pm0.00254$ & 3591.79201 & $-3.19$ \\
 19  &    \nodata                   & 3599.77564 & \nodata & 19.45345& $3606.40434 \pm0.00371$ & 3606.40057 & $5.42$ \\
 20  &    $3614.38474\pm   0.00030$ & 3614.38415 & $0.85$  & 20.45345& $3621.01281 \pm0.00399$ & 3621.00911 & $5.32$ \\
 21  &    $3628.99361\pm   0.00038$ & 3628.99271 & $1.31$  & 21.45345& $3635.61695 \pm0.00322$ & 3635.61762 & $-0.97$ \\
 22  &    $3643.60114\pm   0.00036$ & 3643.60131 & $-0.23$ & 22.45345& $3650.22268 \pm0.00518$ & 3650.22613 & $-4.97$ \\
\enddata
\tablenotetext{a}{BJD - 2,455,000}
\tablenotetext{b}{Observed time minus model time in minutes.}
\end{deluxetable}

\clearpage

%
%
\begin{deluxetable}{ccccc}
\tablecaption{Photometric and spectroscopic parameters of the system prior to the photo-dynamical solution\label{tab:pre_ELC}}
\tablewidth{0pt}
\tablehead{
\colhead{Parameter} &
\colhead{Value} &
\colhead{Uncertainty} &
\colhead{Unit} &
\colhead{Source}
}
\startdata
RV semi-amplitude, $K_1$ & 21.619 & 0.007 & ${\rm km~s^{-1}}$ & Martin et al. (2019) \\
Gravity of Primary, $\log g_1$ & 4.0 & 0.08 & cgs & Spectroscopy, this work \\
Metallicity of Primary, [Fe/H]$_1$ & 0.01 & 0.05 & dex & Spectroscopy, this work \\
Projected Rotational Velocity of Primary, $v \sin i$ & 3.6 & 0.6 & ${\rm km~s^{-1}}$ & Spectroscopy, this work \\
Reddening, E(B-V) & 0.02 & 0.01 & mag & Gaia + Photometry, this work \\
Effective Temperature of Primary, ${T_{\rm eff, 1}}$ & 6050 & 80 & K & Spectroscopy, this work \\
Effective Temperature of Primary, ${T_{\rm eff, 1}}$ & 5990 & 110 & K & Gaia + Photometry, this work \\
Radius of Primary, $R_1$ & 1.345 & 0.046 & $R_{\odot}$ & 
Gaia+Photometry, this work\\
Age & 4.4 & 0.2 & Gyr & This work\\
\enddata
\end{deluxetable}

\begin{deluxetable}{lrrrrr}
\tablecaption{Fitted Parameters for \target\label{tab:fitted}}
\tablewidth{0pt}
\tablehead{
\colhead{Parameter\tablenotemark{a}}   & 
\colhead{Best}  &
\colhead{Mode}    &
\colhead{Median} & 
\colhead{$+1\sigma$}  &
\colhead{$-1\sigma$}  
}
\startdata
$T_{\rm conj,bin}$ &  $3336.8245$ & $3336.8237$ & $3336.8242$ & $0.0025$  & $0.0023$  \\
$P_{\rm bin}$ (d) &  $ 14.608561$ & $ 14.608559$ & $ 14.608559$ & $0.000013$  & $0.000012$  \\
$\sqrt{e_{\rm bin}}\cos\omega_{\rm bin}$  &  $-0.18275$ & $-0.18270$ & $-0.18272$ & $0.00040$  & $0.00040$  \\
$\sqrt{e_{\rm bin}}\sin\omega_{\rm bin}$  &  $0.35015$ & $0.35008$ & $0.35020$ & $0.00036$  & $0.00035$  \\
$M_1$ ($M_{\odot}$)  &  $1.038$ & $1.149$ & $1.127$ & $0.068$  & $0.069$  \\
$Q\equiv M_2/M_1$  &  $0.2865$ & $0.2752$ & $0.2774$ & $0.0069$  & $0.0062$  \\
$i$ (deg)  &  $89.658$ & $89.649$ & $89.696$ & $0.178$  & $0.114$  \\
$T_{\rm eff,1}$ (K)  &  $5990.7$ & $6072.4$ & $6040.8$ & $ 98.1$  & $ 91.5$  \\
$T_{\rm eff,2}/T_{\rm eff,1}$   &  $0.5537$ & $0.5548$ & $0.5516$ & $0.0042$  & $0.0047$  \\
$R_1$ ($R_{\odot}$)   &  $1.299$ & $1.338$ & $1.331$ & $0.024$  & $0.026$  \\
$R_1/R_2$    &  $4.307$ & $4.310$ & $4.308$ & $0.013$  & $0.014$  \\
$q_{1,1}$    &  $0.255$ & $0.245$ & $0.255$ & $0.048$  & $0.043$  \\
$q_{2,1}$    &  $0.308$ & $0.299$ & $0.310$ & $0.058$  & $0.050$  \\
$\zeta_p$ (d)    &  $3436.57$ & $3437.26$ & $3437.32$ & $0.83$  & $0.78$  \\
$\zeta_m$  (d)  &  $-3246.29$ & $-3246.84$ & $-3246.97$ & $0.71$  & $0.77$  \\
$\sqrt{e_{\rm pl}}\cos\omega_{\rm pl}$  &  $-0.035$ & $-0.051$ & $-0.058$ & $0.018$  & $0.024$  \\
$\sqrt{e_{\rm pl}}\sin\omega_{\rm pl}$  &  $-0.303$ & $-0.285$ & $-0.290$ & $0.009$  & $0.011$  \\
$i_{\rm pl}$ (deg)  &  $89.22$ & $89.29$ & $89.37$ & $0.35$  & $0.00$  \\
$\Omega_{\rm pl}$ (deg)  &  $ 0.87$ & $ 0.83$ & $ 0.91$ & $0.35$  & $0.35$  \\
$M_3$ ($M_{\oplus}$)  &  $30.2$ & $33.2$ & $33.0$ & $20.3$  & $19.6$  \\
$R_1/R_3$   &  $21.12$ & $21.20$ & $21.17$ & $0.47$  & $0.44$  \\
$\gamma_1$\tablenotemark{b} (km s$^{-1}$)   &  $30.74769$ & $30.74766$ & $30.74778$ & $0.00089$  & $0.00084$  \\
$\gamma_2$\tablenotemark{c} (km s$^{-1}$)   &  $30.74621$ & $30.74618$ & $30.74641$ & $0.00128$  & $0.00125$  \\
$\gamma_3$\tablenotemark{d} (km s$^{-1}$)   &  $30.75866$ & $30.75836$ & $30.75857$ & $0.00098$  & $0.00098$  \\
\enddata
\tablenotetext{a}{Osculating parameters valid at BJD 2,455,186.0000.}
\tablenotetext{b}{Relative velocity offset, ``early'' CORALIE data.}
\tablenotetext{c}{Relative velocity offset, ``late'' CORALIE data.}
\tablenotetext{d}{Relative velocity offset, HARPS data.}
\end{deluxetable}

\clearpage

%
%
\begin{deluxetable}{lrrrrr}
\tablecaption{Derived Parameters for \target\label{tab:derived}}
\tablewidth{0pt}
\tablehead{
\colhead{Parameter\tablenotemark{a}}   & 
\colhead{Best}  &
\colhead{Mode}    &
\colhead{Median} & 
\colhead{$+1\sigma$}  &
\colhead{$-1\sigma$}  
}
\startdata
\multicolumn{6}{c}{{Bulk Properties}} \\
\hline
$M_1$ ($M_{\odot}$) &  $1.038$ & $1.149$ & $1.127$ & $0.068$  & $0.069$  \\
$R_1$ ($R_{\odot}$) &  $1.299$ & $1.338$ & $1.331$ & $0.024$  & $0.026$  \\
$M_2$ ($M_{\odot}$)  &  $0.2974$ & $0.3168$ & $0.3128$ & $0.0113$  & $0.0118$  \\
$R_2$ ($R_{\odot}$)  &  $0.3015$ & $0.3102$ & $0.3089$ & $0.0056$  & $0.0060$  \\
$M_3$ ($M_{\oplus}$)  &  $30.2$ & $33.2$ & $33.0$ & $20.3$  & $19.6$  \\
$R_3$ ($R_{\oplus}$)  &  $6.71$ & $6.83$ & $6.85$ & $0.19$  & $0.19$  \\
$\rho_3$ (g cm$^{-3}$)  &  $0.55$ & $0.57$ & $0.56$ & $0.34$  & $0.33$  \\
\hline
\multicolumn{6}{c}{{Binary Orbit}} \\
\hline
$P_{\rm bin}$ (d)  &  $14.608561$ & $14.608559$ & $14.608559$ & $0.000013$  & $0.000012$  \\
$T_{\rm conj,bin}$   &  $3336.8245$ & $3336.8237$ & $3336.8242$ & $0.0025$  & $0.0023$  \\
$K_{\rm bin}$ (km s$^{-1}$)   &  $21.6248$ & $21.6246$ & $21.6247$ & $0.0034$  & $0.0032$  \\
$e_{\rm bin}$    &  $0.15601$ & $0.15603$ & $0.15603$ & $0.00015$  & $0.00015$  \\
$\omega_{\rm bin}$  (deg)   &  $117.561$ & $117.568$ & $117.554$ & $0.072$  & $0.074$  \\
$a_{\rm bin}$ (AU)    &  $0.1288$ & $0.1319$ & $0.1321$ & $0.0024$  & $0.0025$  \\
true anomaly (deg) &  $111.217$ & $111.246$ & $111.226$ & $0.071$  & $0.069$  \\
mean anomaly (deg) &  $ 93.882$ & $ 93.897$ & $ 93.889$ & $0.065$  & $0.065$  \\
mean longitude (deg)  &  $228.779$ & $228.783$ & $228.779$ & $0.020$  & $0.020$  \\
$i_{\rm bin}$ (deg)    &  $89.658$ & $89.649$ & $89.696$ & $0.178$  & $0.114$  \\
$\Omega_{\rm bin}$  (deg)  &  $0.0$ & $0.0$ & $0.0$ & $ 0.0$  & $0.0$  \\
\hline
\multicolumn{6}{c}{{Planet Orbit}} \\
\hline
$P_{\rm pl}$ (d)  &  $95.141$ & $95.175$ & $95.174$ & $0.031$  & $0.035$  \\
$T_{\rm conj,pl}$  &  $3341.43$ & $3342.11$ & $3342.15$ & $0.80$  & $0.74$  \\
$e_{\rm pl}$   &  $0.0928$ & $0.0861$ & $0.0880$ & $0.0043$  & $0.0033$  \\
$\omega_{\rm pl}$ (deg)  &  $263.3$ & $260.3$ & $258.6$ & $3.7$  & $4.8$  \\
$a_{\rm pl}$ (AU)  &  $0.4491$ & $0.4639$ & $0.4607$ & $0.0084$  & $0.0088$  \\
true anomaly (deg)  &  $136.0$ & $141.0$ & $141.7$ & $5.2$  & $4.5$  \\
mean anomaly (deg)  &  $128.3$ & $134.1$ & $135.2$ & $6.1$  & $5.4$  \\
mean longitude (deg)  &  $400.2$ & $401.6$ & $401.2$ & $0.9$  & $1.1$  \\
$i_{\rm pl}$ (deg)   &  $89.22$ & $89.29$ & $89.37$ & $0.35$  & $0.26$  \\
$\Omega_{\rm pl}$ (deg)   &  $0.87$ & $0.83$ & $0.91$ & $0.35$  & $0.35$  \\
$I$\tablenotemark{b} (deg)   &  $0.98$ & $0.92$ & $0.99$ & $0.31$  & $0.28$  \\
\enddata
\tablenotetext{a}{Osculating parameters valid at BJD 2,455,186.0000.}
\tablenotetext{b}{Mutual inclination between orbital planes.}
\end{deluxetable}

\begin{deluxetable}{rrrr}
\tablecolumns{4}
\tabletypesize{\footnotesize}
\tablecaption{Initial Dynamical Parameters\tablenotemark{a}\label{tab:Cartesian}}
\tablewidth{0pt}
\tablehead{  
\colhead{parameter\tablenotemark{b}} & 
\colhead{binary orbit}   & 
\colhead{planet orbit}   & 
\colhead{} }
\startdata
Period  (days)&$ 1.46085607280931704E+01$ & $ 9.51407742682822573E+01$ & \cr 
$e\cos\omega$ &$-7.21833507304852212E-02$ & $-1.07723606253091915E-02$ & \cr 
$e\sin\omega$ &$ 1.38301268444568498E-01$ & $-9.22020142859296676E-02$ & \cr 
$i$  (rad) &$ 1.56482063218343082E+00$ & $ 1.55715854420219579E+00$ & \cr 
$\Omega$  (rad) &$ 0.00000000000000000E+00$ & $ 1.52632127894949884E-02$ & \cr 
$T_{\rm conj}$  (days)\tablenotemark{c} &$ 3.33682449014051008E+03$ & $ 3.34142761690665247E+03$ & \cr 
$a$  (AU) &$ 1.28783121829547487E-01$ & $ 4.49132971740733966E-01$ & \cr 
$e$     &$ 1.56005374830665650E-01$ & $ 9.28291720948987292E-02$ & \cr 
$\omega$  (deg) &$ 1.17561331987513597E+02$ & $ 2.63336097695842795E+02$ & \cr 
true anomaly  (deg) &$ 1.11217395295404202E+02$ & $ 1.36038415640448932E+02$ & \cr 
mean anomaly  (deg) &$ 9.38816596366525289E+01$ & $ 1.28272133285056185E+02$ & \cr 
mean longitude  (deg) &$ 2.28778727282917799E+02$ & $ 4.00249031010939916E+02$ & \cr 
$i$  (deg) &$ 8.96576179191039415E+01$ & $ 8.92186126155212662E+01$ & \cr 
$\Omega$  (deg) &$ 0.00000000000000000E+00$ & $ 8.74517674648163101E-01$ & \cr 
\noalign{\vskip 2mm}\hline\noalign{\vskip 2mm}
\multicolumn{1}{c}{parameter\tablenotemark{d}} & \multicolumn{1}{c}{body 1} & 
\multicolumn{1}{c}{body 2} & \multicolumn{1}{c}{body 3} \cr
\noalign{\vskip 2mm}\hline\noalign{\vskip 2mm}
Mass ($M_{\odot}$) & $ 1.03784970719363567E+00$ & $ 2.97388770751337850E-01$ & $ 9.06017229632760055E-05$ \cr 
$x$ (AU) & $ 1.95196590778876217E-02$ & $-6.82335050435495666E-02$ & $ 3.68709659800795730E-01$ \cr 
$y$ (AU) & $ 1.32648896371446897E-04$ & $-4.65900424342245798E-04$ & $ 9.75628498103004414E-03$ \cr 
$z$ (AU) & $ 2.22880532978689296E-02$ & $-7.78747286204819755E-02$ & $ 3.02645766911772141E-01$ \cr 
$v_x$ (AU day$^{-1}$) & $-7.66586168788931707E-03$ & $ 2.67578221143398472E-02$ & $-1.61533175726252566E-02$ \cr 
$v_y$ (AU day$^{-1}$) & $ 5.45640776560390306E-05$ & $-1.90441050119945319E-04$ & $ 6.31089229491667655E-05$ \cr 
$v_z$ (AU day$^{-1}$) & $ 9.13006908476301365E-03$ & $-3.18697187907663951E-02$ & $ 2.27034214206392367E-02$ \cr 
\enddata
\tablenotetext{a}{Reference time =       186.00000,
integration step size = 0.05000 days}
\tablenotetext{b}{Jacobian instantaneous (Keplerian) elements}
\tablenotetext{c}{Times are relative to BJD 2,455,000.000}
\tablenotetext{d}{Barycentric Cartesian coordinates}
\end{deluxetable}

\begin{deluxetable}{lccccccc}
\tabletypesize{\small}
\tablecaption{Times, durations, and impact parameters 
of future planet transits\label{tab:future}}
\tablewidth{0pt}
\tablehead{
\colhead{ BJD - 2,455,000}  & 
\colhead{Year}  &
\colhead{Month}  &
\colhead{Day}  &
\colhead{UTC}  &
\colhead{Impact}  &
\colhead{Duration} &
\colhead{Transit} \\
\colhead{}  & 
\colhead{}  &
\colhead{}  &
\colhead{}  &
\colhead{}  &
\colhead{parameter}  &
\colhead{(hr)} &
\colhead{fraction}
 }
\startdata
$ 3861.1845\pm 0.0161$ & 2020 & Jan& 12  & 16:25:40.7 & $ 0.143\pm 0.128$ & $ 6.74\pm 0.16$   & 100.0\%  \\ 
$ 3954.1353\pm 0.0214$ & 2020 & Apr& 14  & 15:14:50.2 & $ 0.206\pm 0.108$ & $ 9.80\pm 0.23$   & 100.0\%  \\ 
$ 4049.6930\pm 0.0196$ & 2020 & Jul& 19  & 04:37:53.2 & $-0.006\pm 0.189$ & $ 7.91\pm 0.25$   & 100.0\%  \\ 
$ 4142.3991\pm 0.0224$ & 2020 & Oct& 19  & 21:34:40.9 & $ 0.055\pm 0.166$ & $ 7.46\pm 0.18$   & 100.0\%  \\ 
$ 4237.9220\pm 0.0236$ & 2021 & Jan& 23  & 10:07:40.6 & $-0.131\pm 0.236$ & $ 9.73\pm 0.63$   &  99.9\%  \\ 
$ 4330.9984\pm 0.0264$ & 2021 & Apr& 26  & 11:57:42.7 & $-0.128\pm 0.232$ & $ 6.31\pm 0.38$   &  99.9\%  \\ 
$ 4425.7194\pm 0.0319$ & 2021 & Jul& 30  & 05:15:54.7 & $-0.224\pm 0.266$ & $11.78\pm 1.20$   &  99.6\%  \\ 
$ 4519.6951\pm 0.0328$ & 2021 & Nov&  1  & 04:40:55.7 & $-0.315\pm 0.285$ & $ 5.94\pm 0.79$   &  98.7\%  \\ 
$ 4708.3278\pm 0.0389$ & 2022 & May&  8  & 19:52:05.0 & $-0.468\pm 0.308$ & $ 6.06\pm 1.27$   &  92.9\%  \\ 
$ 4801.0174\pm 0.0461$ & 2022 & Aug&  9  & 12:24:59.3 & $-0.417\pm 0.304$ & $ 8.44\pm 1.58$   &  95.5\%  \\ 
$ 4896.7585\pm 0.0461$ & 2022 & Nov& 13  & 06:12:14.7 & $-0.566\pm 0.310$ & $ 6.78\pm 1.72$   &  82.4\%  \\ 
$ 4989.3540\pm 0.0509$ & 2023 & Feb& 13  & 20:29:49.0 & $-0.540\pm 0.313$ & $ 6.04\pm 1.46$   &  85.2\%  \\ 
$ 5084.8165\pm 0.0693$ & 2023 & May& 20  & 07:35:48.3 & $-0.624\pm 0.310$ & $ 8.19\pm 2.34$   &  72.6\%  \\ 
$ 5177.9790\pm 0.0600$ & 2023 & Aug& 21  & 11:29:44.2 & $-0.633\pm 0.317$ & $ 4.93\pm 1.45$   &  70.0\%  \\ 
$ 5272.3566\pm 0.1296$ & 2023 & Nov& 23  & 20:33:30.0 & $-0.650\pm 0.311$ & $ 9.79\pm 2.97$   &  64.5\%  \\ 
$ 5366.6770\pm 0.0742$ & 2024 & Feb& 26  & 04:14:52.4 & $-0.670\pm 0.318$ & $ 4.79\pm 1.52$   &  54.2\%  \\ 
$ 5459.7304\pm 0.1788$ & 2024 & May& 29  & 05:31:47.0 & $-0.665\pm 0.315$ & $ 8.76\pm 2.76$   &  55.7\%  \\ 
$ 5555.2836\pm 0.0967$ & 2024 & Sep&  1  & 18:48:23.7 & $-0.690\pm 0.320$ & $ 5.26\pm 1.77$   &  42.9\%  \\ 
$ 5647.6474\pm 0.1677$ & 2024 & Dec&  3  & 03:32:17.7 & $-0.687\pm 0.322$ & $ 6.27\pm 2.08$   &  44.5\%  \\ 
$ 5743.6224\pm 0.1383$ & 2025 & Mar&  9  & 02:56:12.7 & $-0.707\pm 0.320$ & $ 6.35\pm 2.21$   &  35.5\%  \\ 
$ 5836.0445\pm 0.1543$ & 2025 & Jun&  9  & 13:04:05.3 & $-0.700\pm 0.329$ & $ 4.93\pm 1.67$   &  34.2\%  \\ 
$ 5931.4246\pm 0.2222$ & 2025 & Sep& 12  & 22:11:25.1 & $-0.713\pm 0.320$ & $ 8.39\pm 3.01$   &  31.0\%  \\ 
$ 6024.6162\pm 0.1615$ & 2025 & Dec& 15  & 02:47:16.6 & $-0.694\pm 0.333$ & $ 4.71\pm 1.64$   &  27.7\%  \\ 
$ 6118.5043\pm 0.3348$ & 2026 & Mar& 19  & 00:06:09.6 & $-0.702\pm 0.325$ & $ 9.86\pm 3.45$   &  27.4\%  \\ 
$ 6213.1325\pm 0.1991$ & 2026 & Jun& 21  & 15:10:46.2 & $-0.687\pm 0.333$ & $ 5.22\pm 1.81$   &  23.9\%  \\ 
$ 6305.7169\pm 0.3200$ & 2026 & Sep& 22  & 05:12:19.8 & $-0.698\pm 0.333$ & $ 7.40\pm 2.63$   &  23.8\%  \\ 
$ 6401.3824\pm 0.2847$ & 2026 & Dec& 26  & 21:10:36.5 & $-0.685\pm 0.332$ & $ 6.52\pm 2.34$   &  22.4\%  \\ 
$ 6493.7147\pm 0.2750$ & 2027 & Mar& 29  & 05:09:06.7 & $-0.673\pm 0.340$ & $ 5.53\pm 1.88$   &  20.7\%  \\ 
$ 6589.0498\pm 0.4693$ & 2027 & Jul&  2  & 13:11:42.7 & $-0.678\pm 0.329$ & $ 8.97\pm 3.27$   &  21.5\%  \\ 
$ 6682.0668\pm 0.2875$ & 2027 & Oct&  3  & 13:36:11.6 & $-0.665\pm 0.342$ & $ 5.11\pm 1.70$   &  19.6\%  \\ 
$ 6775.8467\pm 0.7051$ & 2028 & Jan&  5  & 08:19:17.2 & $-0.675\pm 0.332$ & $10.95\pm 3.69$   &  20.4\%  \\ 
$ 6870.3964\pm 0.3605$ & 2028 & Apr&  8  & 21:30:48.3 & $-0.672\pm 0.345$ & $ 5.65\pm 1.98$   &  20.7\%  \\ 
$ 6962.8655\pm 0.5975$ & 2028 & Jul& 10  & 08:46:15.3 & $-0.666\pm 0.340$ & $ 7.74\pm 2.61$   &  19.6\%  \\ 
$ 7058.3711\pm 0.5322$ & 2028 & Oct& 13  & 20:54:25.6 & $-0.672\pm 0.341$ & $ 7.42\pm 2.70$   &  23.1\%  \\ 
$ 7150.8050\pm 0.4580$ & 2029 & Jan& 14  & 07:19:14.6 & $-0.661\pm 0.351$ & $ 5.64\pm 1.88$   &  21.1\%  \\ 
$ 7245.5142\pm 0.8968$ & 2029 & Apr& 19  & 00:20:24.3 & $-0.682\pm 0.341$ & $10.53\pm 4.07$   &  26.0\%  \\ 
$ 7339.0599\pm 0.4556$ & 2029 & Jul& 21  & 13:26:12.1 & $-0.664\pm 0.361$ & $ 5.33\pm 1.87$   &  25.6\%  \\ 
$ 7431.8295\pm 1.0387$ & 2029 & Oct& 22  & 07:54:30.9 & $-0.685\pm 0.351$ & $10.20\pm 3.65$   &  27.5\%  \\ 
$ 7527.1701\pm 0.5751$ & 2030 & Jan& 25  & 16:04:55.6 & $-0.696\pm 0.361$ & $ 6.19\pm 2.34$   &  36.4\%  \\ 
\enddata
\end{deluxetable}



\bibliography{bibfile}{}
\bibliographystyle{aasjournal}

\end{document}